\documentclass[aps,prd,twocolumn,superscriptaddress,nofootinbib,floatfix]{revtex4-2}
\usepackage{amsmath}
\usepackage{cases}
\usepackage{amsfonts}
\usepackage{amssymb} 
\usepackage{graphicx}
\usepackage{float}
\usepackage{xcolor}
\usepackage{dsfont}
\DeclareUnicodeCharacter{2212}{-}

\newcommand{\K}{{\cal K}}

\begin{document}

\title{Two-fluid stellar objects in General Relativity: the covariant formulation}
\author{Nolene F. Naidu}
\email[]{nolene.naidu@physics.org}
\affiliation{Department of Mathematics and Applied Mathematics, University of Cape Town}
\author{Sante Carloni}
\email[]{sante.carloni@unige.it}
\affiliation{DIME Sez. Metodi e Modelli Matematici, Universit\`{a} di Genova,\\ Via All'Opera Pia 15, 16145 - Genoa, (Italy).}
\author{Peter Dunsby}
\affiliation{Department of Mathematics and Applied Mathematics, University of Cape Town}
\date{\today}

\begin{abstract}
We apply the 1+1+2 covariant approach to describe a general static and spherically symmetric relativistic stellar object which contains two interacting fluids. We then use the 1+1+2 equations to derive the corresponding Tolman-Oppenheimer-Volkoff (TOV) equations in covariant form in the isotropic, non-interacting case. These equations are used to obtain new exact solutions by means of direct resolution and reconstruction techniques. Finally, we show that the generating theorem known for the single fluid case can also be used to obtain two-fluid solutions from single fluid ones. 
\end{abstract}

\maketitle

\section{Introduction}
The derivation of solutions for relativistic stellar objects is a notoriously complex problem when approached using analytical methods. It is probably for this reason that most of the attempts to find such solutions have up to now relied on modeling the matter distribution using a single fluid description. In realistic situations, however, such an assumption is not physically realistic for several reasons: compact stars stars can have a structure that can be wildly different in terms of composition and pressure, and there could be interactions between different components. Hence, in order to build more accurate models, it is necessary to turn to a multi-fluid description of these objects.

So far, there have been several attempts to model stellar objects containing different fluids. The majority of these rely on using numerical, non covariant approaches (see e.g. \cite{numer}). A different perspective was provided for the first time by Carter and Langlois \cite{carter}. They showed that it is possible to formulate a covariant exact model of multi-fluid (neutron) stars using non-interacting fluids and assuming an equation of state. Although in the end the analysis of the equations in \cite{carter} is still numeric, their attempt shows that a covariant approach to the TOV equations might be useful to uncover new aspects of these equations.

The present work aims to construct a different approach, which makes full use of covariance and is oriented towards an analytical investigation of the TOV equations. There are several reasons why it is important to develop analytical studies of the TOV equations in parallel to numerical studies. For example, exact solutions can be used to explore the full parameter space for a given metric, rather than a single set of values. In addition, exact solutions can be used to test numerical codes, particularly when they entail new languages/approximations schemes.    

We will also show that our analytical approach is able to include other equations of state, can be generalized to any number of fluids, fields, and can include fluxes and interactions.

The cornerstone of our formulation will be the Tolman-Oppenheimer-Volkoff (TOV) equations of hydrostatic equilibrium. These equations were introduced in 1939 \cite{tolman} \cite{oppvolk}, and provide insights into the pressure profile of a static, spherically symmetric object in General Relativity (GR). Since their introduction, several authors have tried to solve these equations exactly (see, e.g. \cite{del} for a list of exact solutions). In spite of these efforts, the resolution of the TOV equations still remains a formidable task, particularly if one aims at the deduction of realistic solutions.

In two recent papers, the TOV equations were presented in a fully covariant form and applied to the case of isotropic and anisotropic fluids \cite{sante1,sante2,Luz:2019frs}. The generalized equations are written in a covariant dimensionless and autonomous form, thereby providing a combination of the Lane-Emden and homology invariant formulation \cite{Kimura,Horedt}. The covariant (and therefore observer independent) form of the equations presents many benefits. For example, one may change coordinate systems with ease, making the description of the system (i.e. its symmetries and properties) easily captured.

The covariant formulation of the TOV equations is based on the so-called {\it covariant approach}. The original development of the covariant formalisms is due to Ehlers, Ellis and other authors \cite{ehlers}. Their {\it 1+3 covariant approach} offers a powerful method for studying the general properties of exact relativistic (and Newtonian) cosmological models \cite{Cargese}. Because all the gravitational and fluid equations can be written down exactly, this approach is well suited to a top-down construction of perturbation theory and has been widely used in studies of perturbations of Friedmann-Robertson-Walker models \cite{FLRW-pert} and other backgrounds that admit a high degree of symmetry.

An extension of the 1+3 covariant approach, known as the {\it 1+1+2 covariant approach}, proposed by Greenberg \cite{green} and adapted to Locally Rotationally Symmetric (LRS) spacetimes by Clarkson \cite{clarkbar}, \cite{clarkson} and Betschart \cite{betschart} allows for the application of the formalism to numerous astrophysical scenarios (for example, lensing in spherically symmetric spacetimes \cite{nzioki}, or spherically symmetric spacetimes \cite{Carloni:2014rba}). The further "split" or "foliation" is in a spatial direction, leaving the other two spatial dimensions unchanged.

The 1+1+2 TOV equations have already proven to offer a useful new perspective on the problem of finding exact models for relativistic stars. They are generally easier to solve than the original ones and allow a direct application of reconstruction methods. In addition, in \cite{sante1,sante2} it was shown that the generating theorems proposed by Boonserm {\it et al} in \cite{boon1, boon2} can be easily formulated and extended to more complex cases. The 1+1+2 TOV equations have also been used to uncover new features of relativistic stellar objects in Einstein-Cartan gravity in \cite{Luz:2019frs}.

In the following, we will provide a complete description of the interior of a relativistic star composed of two fluids using the 1+1+2 formalism. In particular, we will use the Ricci identities, the Bianchi Identities, and the Einstein equations to derive a complete system of covariant TOV equations for two fluids with non-zero fluxes and can be generalized to include anisotropies and interactions. Then we will explore some exact solutions which are physically relevant according to the criteria given in \cite{del}. We will also extend the generating theorems, proving that they can be used to obtain two-fluid solutions from single fluid ones.  

The outline of this paper is as follows: In Sec. \ref{TFC} the 1+1+2 equations in the case of two fluids are used to construct the covariant TOV equations in the isotropic and non-interacting case. Section \ref{CON} deals with the conditions of physical viability for a given solution of the TOV equations. Section \ref{Junction}, instead, gives a brief sketch of the generalization of Israel's junction conditions to the multi-fluid case. Section \ref{KS} contains information on some known solutions which will be useful to obtain the main results of the paper. In Sec. \ref{CD} we obtain the two-fluid generalization of the interior Schwarzschild solution. In Sec. \ref{SR} we employ a reconstruction algorithm to derive new two-fluid solutions based on the single fluid ones of Sec. \ref{KS}. The two-fluid extension of the generating theorems of \cite{boon1} is discussed in Sec. \ref{GT}. In Sec. \ref{SecQ} we consider some exact two-fluid solutions which include (stationary) fluxes. Finally, a discussion and some concluding remarks can be found in Sec. \ref{CR}. The main equations of the 1+1+2 formalism are presented in Appendix \ref{AA} and the $N$ fluid generalization of the 1+1+2 equations which include anisotropic pressure and interactions in Appendix \ref{App2}.
\section{The 1+1+2 equations for the two-fluid case}\label{TFC}
Coordinate invariant and tetrad methods are an important way of transforming the equations of General Relativity into first order ODEs, as opposed to second order PDEs. The approach is most useful in the presence of homogeneity, isotropy, and spacetimes that admit a high degree of symmetry. The 1+1+2 formalism that we will employ in the following can be considered a semi-tetrad approach because it relies on both a time-like and a space-like threading\footnote{To be precise, we should point out that the 1+1+2 formalism is somewhat less general than the tetrad one: the former assumes that the vector fields used for the threading are everywhere regular.}.  

We, therefore, start constructing the 1+1+2 formalism from the threading decomposition of the spacetime. In this way, we can construct a set of tensorial objects connected to the properties of the field lines, which make up the set of $1+1+2$ variables. Following this, we use the Bianchi and Ricci identities, together with the Einstein equations, to derive a closed system of first order propagation and constraint equations. In this section, we present the equations for the two-fluid case and refer the reader to Appendix \ref{AA} for details of the general equations and formalism. 

As it can be seen in the Appendix \ref{App2}, the 1+1+2 equations can be written down easily for any number of interacting fluids. However, in this work we will limit ourselves to two non interacting fluids. Considering two fluids is justified, other than by simplicity, by the fact that two fluids models are already enough to describe systems like neutron stars, which are one of the main applications of the TOV equations \cite{lang}. 

We define a time-like threading vector field $u^a$ associated to the observer's congruence with $u^a u_a = -1$, and a space-like vector $e_a$ with $e_a e^a = 1$. The $u^a$ and $e_a$ congruences describe a geometry defined by two projection tensors given by
\begin{eqnarray}
{h^a}_b &=& {g^a}_b + u^a u_b \hspace{0.2cm},\hspace{0.2cm} {h^a}_a = 3, \nonumber \\
{N_a}^b &=& {h_a}^b - e_a e^b = {g_a}^b + u_a u^b - e_a e^b \hspace{0.1cm},\hspace{0.1cm}{N^a}_a = 2,\hspace{0.6cm}
\end{eqnarray}
where ${h^a}_b$ represents the metric of the 3-spaces orthogonal to $u_a$, and ${N_a}^b$ represents the metric of the 2-spaces orthogonal to $u_a$ and $e_a$. Any tensorial object may now be split according to the above foliations \cite{sante1}. The covariant time derivative, orthogonally projected covariant derivative, hat-derivative and $\delta$-derivative are given by
\begin{eqnarray}
{\dot{X}{}^{a..b}}_{c..d} &\equiv& u^e {\nabla}_e {X^{a..b}}_{c..d}, \nonumber \\
D_e {X^{a..b}}_{c..d} &\equiv& {h^a}_f...{h^b}_g{h^p}_c... {h^q}_d {h^r}_e {\nabla}_r {X^{f..g}}_{p..q}, \nonumber \\
{\hat{X}{}_{a..b}}^{c..d} &\equiv& e^f D_f {X_{a..b}}^{c..d}, \nonumber \\
{\delta}_e {X_{a..b}}^{c..d} &\equiv& {N_a}^f...{N_b}^g {N_i}^c...{N_j}^d {N_e}^p D_p {X_{f..g}}^{i..j}. \hspace{0.3cm}
\end{eqnarray}
In the following we will consider only LRSII spacetimes. The kinematical variables which we will employ are given by (see Appendix \ref{AA} for the complete list of variables for a general LRS spacetime)
\begin{subequations} 
\begin{align} 
& \mathcal{A} = e_a \dot{u}^a,  \\
&\phi = {\delta}_a e^a,  \\
& \mathcal{E} = {C^{ac}}_{bd} u_c u^d \left(e_a e^b - \frac{1}{2} N_{a}^b \right),  
\end{align}
\end{subequations} 
where $C_{abcd}$ is the Weyl tensor. 

The energy-momentum tensor is decomposed as
\begin{eqnarray}
T_{ab} = &&\mu u_a u_b + p\left(e_a e_b+N_{ab}\right) +2 Q e_{(a} u_{b)} .
\end{eqnarray}
The matter variables are given by
\begin{eqnarray}
\mu &=& T_{ab} u^a u^b, \nonumber \\
p &=& \frac{1}{3} T_{ab} \left(e^a e^b + N^{ab}\right),  \\
Q &=& \frac{1}{2} T_{ab} e^a u^b, \nonumber
\end{eqnarray}
where $\mu$ is the density, $p$ is the pressure, and $Q$ is the scalar component of the heat flux.

Selecting a frame in which the total flux is zero, the system (\ref{hatsystem1}) for two isotropic fluids is given by\footnote{The generalization for equations in the case of $n$ fluids and the presence of anisotropic pressure can be found in Appendix \ref{App2}.}
\begin{subequations} \label{hatsystem2}
\begin{align} 
\hat{\phi}=&-\frac{1}{2} {\phi}^2 - \frac{2}{3} ({\mu}_1 + {\mu}_2) - \mathcal{E} , \\
\hat{\mathcal{E}}=& \frac{1}{3} (\hat{\mu}_1 - \hat{\mu}_2) -\frac{3}{2} {\phi} \mathcal{E}, \\ 
-\mathcal{A} \phi &+ \frac{1}{3}({\mu}_1 + 3 p_1) + \frac{1}{3}({\mu}_2 + 3 p_2)- \mathcal{E} = 0,\\
\hat{\mathcal{A}}=&- \mathcal{A} \left( \mathcal{A} + \phi \right) + \frac{1}{2}({\mu}_1 + 3p_1)\nonumber\\
&+ \frac{1}{2}({\mu}_2 + 3p_2), \\
\hat{p}_1 =&- \mathcal{A} \left( {\mu}_1 + p_1 \right), \\
 \hat{p}_2=&- \mathcal{A} \left( {\mu}_2 + p_2 \right)  ,\\
K=&\frac{1}{3} ({\mu}_1 + {\mu}_2) - \mathcal{E} + \frac{1}{4} {\phi}^2 ,\\
\hat{Q}_1=&-Q_1 \left(\phi + 2 \mathcal{A} \right) , \label{Q1}\\
Q_2 =& - Q_1 , \label{Q2}\\
\hat{K}=&-\phi K. \label{Keq}
\end{align}
\end{subequations}
Next, we introduce a useful parameter, named $\rho$, such that $\hat{X} = \phi X_{,\rho}$. In this way the equation for the Gauss' curvature $K$ \eqref{Keq} can be solved to give
\begin{equation} \label{Krho}
K = K_0^{-1} e^{-\rho}.
\end{equation}
Defining the variables
\begin{align}
X &= \frac{{\phi}_{, \rho}}{\phi}, & Y &= \frac{\mathcal{A}}{\phi}, \nonumber \\
 \mathcal{K} &= \frac{K}{{\phi}^2} , &E &= \frac{\mathcal{E}}{{\phi}^2}, \nonumber \\
\mathbb{M}_1 &= \frac{{\mu}_1}{{\phi}^2}, & \mathbb{M}_2 &= \frac{{\mu}_2}{{\phi}^2}, \label{newvars}\\
P_1 &= \frac{p_1}{{\phi}^2}, & P_2 &= \frac{p_2}{{\phi}^2}, \nonumber \\
\mathbb{Q}_1 &= \frac{Q_1}{{\phi}^2}, &
\mathbb{Q}_2 &= \frac{Q_2}{{\phi}^2} ,\nonumber
\end{align}
and using $\rho$ as parameter, we can recast Eqs. (\ref{hatsystem2}) as
\begin{subequations}
\begin{align}
Y_{, \rho} =& - Y ( X + Y + 1) + \frac{1}{2}(\mathbb{M}_1 + \mathbb{M}_2) \nonumber \\ &+ \frac{3}{2}(P_1 + P_2), \label{DE1} \\
\mathcal{K}_{, \rho} =& -\mathcal{K}(1 + 2 X), \label{DE2} \\
{P}_{1, \rho} = &- Y (\mathbb{M}_1 + P_1) - 2 X P_1, \label{DE3} \\
{P}_{2, \rho} = &- Y (\mathbb{M}_2 + P_2) - 2 X P_2,\\
\mathbb{Q}_{1, \rho}=& -\mathbb{Q}_1 ( 1 + 2 X + 2 Y),
\end{align}
\end{subequations}
with the following constraints
\begin{subequations}\label{constraints}
\begin{eqnarray}
2(\mathbb{M}_1 + \mathbb{M}_2) + 2 (P_1 + P_2) + 2 X - 2 Y + 1 = 0, \label{c1} \\
1 - 4 \mathcal{K} - 4 (P_1 + P_2) + 4 Y = 0, \label{c2} \\
2(\mathbb{M}_1 + \mathbb{M}_2)+ 6 (P_1 + P_2) - 6 Y - 6 E = 0, \label{c3}
\end{eqnarray}
\end{subequations}
and
\begin{eqnarray}\label{constraints2}
E &=& \frac{1}{3} \left( \mathbb{M}_1 + \mathbb{M}_2 \right) + P_1 + P_2 - Y, \\
X &=& -\frac{1}{2} - (\mathbb{M}_1 + \mathbb{M}_2) - (P_1 + P_2) + Y, \\
\mathbb{Q}_{1}&=&-\mathbb{Q}_{2}.
\end{eqnarray}

It is always possible, and sometimes useful to write the equation for the total pressure, which reads
\begin{equation}
{P}_{tot, \rho} + P_{tot}(Y+2X) + Y \mathbb{M}_{tot} =0 ,
\end{equation}
where $P_{tot}=P_{1}+P_2$. The covariant equivalent of the TOV equations can be obtained using the constraints to eliminate all the metric related variables except $\mathcal{K}$. In this way one obtains
\begin{widetext}
\begin{eqnarray}\label{tov2} 
{P}_{1, \rho} &=& - {P_1}^2 + P_1 \left[\mathbb{M}_1 - 3 \mathcal{K} + \frac{7}{4} \right] + \mathbb{M}_1 \left(\frac{1}{4} -\mathcal{K}\right)- P_1 (P_2 - 2\mathbb{M}_2) -\mathbb{M}_1 P_2 , \\
{P}_{2, \rho} &=& - {P_2}^2 + P_2 \left[\mathbb{M}_2 - 3 \mathcal{K} + \frac{7}{4} \right] + \mathbb{M}_2 \left(\frac{1}{4} -\mathcal{K}\right)- P_2 (P_1 - 2\mathbb{M}_1) -\mathbb{M}_2 P_1 , \\
\mathcal{K}_{, \rho} &=& 2 \mathcal{K} \left( \frac{1}{4}-\mathcal{K} + \mathbb{M}_1 + \mathbb{M}_2 \right), \label{K2} \\
\mathbb{Q}_{1, \rho}&=& \mathbb{Q}_1 \left[ 2\mathcal{K} - 2 (\mathbb{M}_1 + \mathbb{M}_2)-\frac{3}{2} \right],\\
{P}_{tot, \rho} &=& - {P_{tot}}^2 + P_{tot} \left[\mathbb{M}_{tot}- 3 \mathcal{K} + \frac{7}{4} \right] + \mathbb{M}_{tot}\left( \frac{1}{4} - \mathcal{K} \right). \label{EqPtot}
\end{eqnarray}
\end{widetext}

In the following we will present some solutions of the above equations for particular cases. It will be useful, then, to give some results which might help the physical interpretation of these solutions. For a generic metric tensor of the form 
\begin{align}\label{metrick4}
ds^2 =& -k_1(x,t) dt^2 + k_2(x,t) dx^2 \nonumber \\
&+ k_3(x,t) \left[dy^2 + k_4(y) dz^2\right],\\
k_4(y) &= \left\{ \begin{array}{ll}
\sin{y},&\mbox{closed geometry}\\
y, &\mbox{flat geometry}\\
\sinh{y},&\mbox{closed geometry}
\end{array}
\right.
\end{align}
we can write \cite{betschart}
\begin{equation}
\begin{split}\label{AphiCoord}
\phi=\frac{\hat{k}_3}{k_3},\quad
\mathcal{A}=\frac{\hat{k}_1}{2k_1}.
\end{split}
\end{equation}
We will use these relations in Sec. \ref{SecQ} where we will deal with models which include fluxes. 

In the same way, in order to give a representation of the solutions obtained below in a form more consistent with the current literature, we give some conversion formulae connecting the 1+1+2 potentials to the parameter $\rho$ and the area radius, $r$. In terms of $\rho$, a generic solution of the TOV equations will be written as 
\begin{equation}\label{metrick}
ds^2 = -k_1(\rho) dt^2 + k_2(\rho) d\rho^2 + k_3(\rho) d\Omega^2,
\end{equation}
where 
\begin{align}
k_3(\rho) &= K_0 e^{\rho},\\
d\Omega^2 &= d\theta^2 + \sin^2 \theta d\phi^2.
\end{align}
In these coordinates we have
\begin{equation} \label{vars}
\begin{split}
\phi&= \frac{1}{\sqrt{k_2}},\quad
\mathcal{A}= \frac{k_{1,\rho}}{2 k_1\sqrt{k_2}},\\
X &= -\frac{k_{2,\rho}}{ k_2} ,\quad
Y = \frac{k_{1,\rho} }{2 k_1}, \\
\mathcal{K} &= \frac{k_2}{K_0 e^{\rho}}, 
\end{split}
\end{equation}
where $K_0$ is a suitable constant. The solutions we find will however be expressed in terms of the area radius $r$ to offer a more familiar representation of our results. The relation between $r$ and $\rho$ is 
\begin{align}
\rho &= 2 \ln \left(\frac{r}{r_0}\right),\\
r &=\sqrt{K_0} e^{\rho/2},
\end{align}
where $r_0$ and $K_0$ are constants related by $K_0=r_0^2$. 

The conversion from the metric coefficient in $\rho$ and the ones in $r$ can be achieved simply by noting that $k_1$ and $k_3$ are scalars with respect to the change of the radial parameter and that 
\begin{equation}
k_2 (\rho)= \frac{r^2}{4}k_2 (r).
\end{equation}
\section{Conditions for physical viability}\label{CON}
Although many solutions can be found to the TOV equations, several of these are not physical. We give here the set of conditions that a solution must satisfy to be physically relevant \cite{del,Carloni:2014rba,sante1,sante2}. Firstly, we require that each fluid satisfy the weak energy condition:
\begin{equation}\label{energycond}
\mu \geq 0, \quad \mu + p \geq 0.
\end{equation}

Secondly, they must satisfy the conditions 
\begin{equation}\label{stabcond}
\mu^{\prime} < 0, \quad p^{\prime} < 0,  
\end{equation}
where the prime represents the derivative with respect to the area radius. These conditions are necessary (but not sufficient) for the stability of the solution.

The third requirement is for causality from the speed of sound
\begin{equation}\label{soundcond}
0< \frac{\partial p}{\partial \mu}<1. 
\end{equation}
We further require that the sources of the Einstein equations are positive definite
\begin{equation}\label{pcond}
p \geq 0.
\end{equation}
Finally, we require that the matter variables are finite and positive valued at the center of the matter distribution. A possible exception to this rule will be discussed in Sec.~\ref{SecQ}.

\section{Junction conditions}\label{Junction} 
Another important aspect of the search for solutions relates to the junction between the interior solution and the exterior vacuum spacetime, characterized by the Schwarzschild metric. The procedure of the determination of the junction conditions is very similar to the case of a single fluid solution treated in \cite{sante1,sante2}. In particular, Israel's junction conditions~\cite{Israel:1966rt,Barrabes:1991ng} are equivalent to
\begin{equation}\label{Israel2}
[\K]=0, \qquad [Y]=0\,.
\end{equation}
Using the constraint in Eq.~\eqref{constraints2} above one gets
\begin{equation}\label{JunCov}
 \left[P_1+P_2\right]=0\,.
\end{equation}
This implies that a smooth junction with the Schwarzschild metric requires the total pressure must be zero at the junction. Since pressures must always be positive definite in a realistic solution, the above result implies that the pressure of both fluids must be zero at the junction. As we shall see, however, the most common occurrence in two-fluid solutions is that one of the pressures goes to zero at a specific value of the radial parameter while the other is not. In this case, one should consider the two-fluid solution up to that distance from the center and match this solution with a single fluid one thereafter. In other words, the inclusion of more than one fluid leads directly to a shelled structure for the matter distribution. As in the case of a single fluid, there is no condition on the energy density and the tangential pressure apart from the ones discussed in the previous section.

In the solution presented below, we will consider vanishing the radial pressure on the vacuum boundary as a desirable feature. Indeed, while there is no need to have a ``hard boundary'' in a stellar object, the inclusion of a ``soft'' boundary would require the introduction of types of sources that we have not included in our treatment i.e. electromagnetic fields, tensions etc. As we have excluded  sources which are not a regular perfect fluid, the request of a hard boundary seems well-motivated from a physical point of view.

\section{Some known solutions for single fluid relativistic stars}\label{KS}
In the following sections, we will consider some indirect resolution methods of the TOV equations \eqref{tov2}. These methods can be implemented in an easier way if we rely on the characteristics of a known single fluid solution. This section aims to introduce three such solutions in a form compatible with the formalism we will employ. We will consider in particular the Interior Schwarzschild \cite{1916skpa.conf.424S}, the Tolman IV \cite{tolman} and Heintzmann IIa \cite{heint} solutions. 

\subsection{Interior Schwarzschild (Constant Density) solution.}

The interior Schwarzschild solution \cite{1916skpa.conf.424S}, was the very first solution for the interior of a static spherically symmetric relativistic object. It assumes the fluid to be incompressible, a feature introduced in the model by assuming a constant density.

For a metric written in the form (\ref{metrick}), the constant density solution is given by
\begin{equation} \label{cdsoln}
\begin{split}
k_1 &= a_0 \left({\mathfrak c}_{1}+z\right)^2, \\
k_2 &= \frac{3}{z^2},\\
k_3 &= r^2,
\end{split}
\end{equation}
where
\begin{equation}
z= \sqrt{3-r^2 \mu_1},
\end{equation}
and $a_0$, ${\mathfrak c}_{1}$ and $\mu_1$ are constants. The metric \eqref{cdsoln} corresponds, via the Einstein equations, to the following expressions for the pressure and energy density respectively
\begin{subequations}\label{cdmv}
\begin{align}
p_{CD}(r) &= -\frac{\mu_1(3 z + {\mathfrak c}_{1})}{3 \left(z + {\mathfrak c}_{1}\right)}, \label{cdp}\\
\mu_{CD} (r) &= \mu_1.
\end{align}
\end{subequations}
Applying the conditions (\ref{energycond}-\ref{pcond}) with the exception of the first equation in (\ref{stabcond}), we obtain with $\mu_1 \neq 0$ the maximum radius for the object described by this metric is:
\begin{align}
r < \frac{1}{3} \sqrt{\frac{27- {{\mathfrak c}_{1}}^2}{\mu_1}}.
\end{align}
which corresponds to the well-known Buchdahl limit \cite{Buch}. In terms of the newly defined variables \eqref{newvars} and the parameter $\rho$, we have 
\begin{equation}
\phi = -\frac{2 z_1}{\sqrt{3K_0 e^{\rho}}},
\end{equation}
and (\ref{cdmv}) correspond to
\begin{subequations}\label{cdnmv}
\begin{align}
\mathcal{K}(\rho) =& \frac{3}{4 {z_1}^2},\\
P(\rho) =& -\frac{\mu_1 K_0 e^{\rho}(3 z_1 + {\mathfrak c}_1)}{4 z_1 ^2 (z_1 + {\mathfrak c}_1)}, \\
\mathbb{M}(\rho) =& \frac{3 \mu_1 K_0 e^{\rho}}{4 z_1 ^2},\\
Y(\rho) =& -\frac{\mu_1 K_0 e^{\rho}}{2 z_1 (z_1 + {\mathfrak c}_1)}, 
\end{align}
\end{subequations}
where
\begin{equation}
z_1 = \sqrt{3-\mu_ 1 K_0 e^{\rho}}.
\end{equation}

\subsection{Tolman IV solution}\label{TolIV}
The Tolman IV solution was presented in the seminal paper by Tolman \cite{tolman} and is a well-known solution with no irregularity at $r=0$. It is characterized by an equation of state that is quadratic in the pressure.

For a metric written in the form (\ref{metrick}), the Tolman IV solution is given by
\begin{equation} \label{tol4}
\begin{split}
k_1 &= B^2 \left(1 + \frac{r^2}{A^2}\right), \\ 
k_2 &= \frac{R^2(A^2 + 2 r^2)}{\left(R^2 - r^2\right) \left(A^2 + r^2 \right)}, \\ 
k_3 &= r^2 ,
\end{split}
\end{equation}
with the following expressions for the pressure and energy density respectively
\begin{subequations}\label{tol4mv}
\begin{align}
p_T(r) &= \frac{R^2-A^2-3 r^2}{R^2 \left(A^2+2 r^2\right)}, \label{tol4p}\\
\mu_T (r) &= \frac{R^2 \left(3 A^2+2 r^2\right)+7 A^2 r^2+3 A^4+6 r^4}{R^2 \left(A^2+2 r^2\right)^2}.
\end{align}
\end{subequations}
Applying the conditions (\ref{energycond}-\ref{pcond}), we find that $A\neq0$ and that the maximum radius for the object described by this metric is:
\begin{align}
&r\leq \sqrt{\frac{R^2-A^2}{3}}.
\end{align}
In terms of the newly defined variables \eqref{newvars} and the parameter $\rho$, we have 
\begin{equation}
\phi_T =-\sqrt{\frac{4 (A^2+ K_0 e^{\rho}) (R^2- K_0 e^{\rho})}{K_0 e^{\rho} R^2 (A^2+2 K_0 e^{\rho})}},
\end{equation}
and (\ref{tol4mv}) correspond to
\begin{subequations}\label{tol4nmv}
\begin{align}
\mathcal{K}_T(\rho) =&\frac{R^2 \left(A^2+2 K_0 e^{\rho}\right)}{4 \left(A^2+ K_0 e^{\rho}\right)
 \left(R^2- K_0 e^{\rho}\right)},\\
P_T(\rho) =& \frac{ K_0 e^{\rho} \left(A^2+3 K_0 e^{\rho}-R^2\right)}{4 \left(A^2+ K_0 e^{\rho}\right) \left( K_0 e^{\rho}-R^2\right)}, \\
\mathbb{M}_T(\rho) =& \frac{A^2 \left(2 A^2+R^2\right)}{4
 \left(A^2+R^2\right) \left(A^2+ K_0 e^{\rho }\right)}-\frac{A^2}{2 \left(A^2+2 K_0 e^{\rho}\right)}
 \nonumber\\
 &+\frac{R^2 \left(3 A^2+4R^2\right)}{4 \left(A^2+R^2\right) \left(R^2- K_0 e^{\rho}\right)}-\frac{3}{4}, \\
Y_T(\rho) =& \frac{K_0 e^{\rho}}{2 \left(A^2+ K_0 e^{\rho}\right)}.
\end{align}
\end{subequations}
\subsection{Heintzmann IIa solution}
The Heintzmann IIa solution was presented for the first time in \cite{heint}. Its metric can be written as
\begin{align} \label{hein2a}
k_1 &= b^2 (1 + a r^2)^3,\\
k_2 &= \left( 1 - \frac{3 a r^2}{2} \frac{1 + c (1 + 4a r^2)^{-\frac{1}{2}}}{1 + a r^2}\right)^{-1}, \\
k_3 &= r^2 ,
\end{align}
with the pressure and energy density given by
\begin{subequations}\label{hein2amv}
\begin{align}
p_H(r) &= -\frac{3 a \left[7 a c r^2+3 \left(a r^2-1\right) \sqrt{4 a r^2+1}+c\right]}{2 \left(a r^2+1\right)^2 \sqrt{4 a r^2+1}}, \\
\mu_H (r) &= \frac{3 a \left[c \left(9 a r^2+3\right)+\left(a r^2+3\right) \left(4 a r^2+1\right)^{3/2}\right]}{2 \left(a r^2+1\right)^2 \left(4 a r^2+1\right)^{3/2}}.
\end{align}
\end{subequations}
The Heintzmann IIa metric satisfies conditions (\ref{energycond}-\ref{pcond}) for $a>0$:
\begin{align}
&r < \frac{1}{\sqrt{a}}.
\end{align}
As before, in order to find some results useful for the next sections we give some of the variables \eqref{newvars} in terms of the parameter $\rho$. We have 
\begin{equation}
\begin{split}
\phi_H^2 =&\frac{4 e^{-\rho }}{K_0}-\frac{6 a}{a K_0
 e^{\rho }+1}\\
 &-\frac{6 a c}{\left(a K_0 e^{\rho }+1\right) \sqrt{4 a K_0 e^{\rho }+1}}.
 \end{split}
\end{equation}
We can write (\ref{hein2amv}) as
\begin{subequations}\label{hein2anmv}
\begin{align}
{\mathcal K}_H(\rho) =&\frac{a K_0 e^{\rho}+1}{4 - 2 a K_0 e^{\rho}\left(1+\frac{3 c}{\sqrt{4 a K_0 e^{\rho}+1}}\right)},\\
P_H(\rho) =& \frac{a K_0 e^{\rho}+1}{ 2 a e^{\rho}K_0 \left(1+\frac{3 c}{\sqrt{4 a K_0 e^{\rho}+1}}\right)-4}\nonumber\\
 &-\frac{3}{2 a K_0 e^{\rho}+2}+\frac{7}{4}, \\
\mathbb{M}_H(\rho) =&\frac{3 a K_0 e^{\rho}}{4 (4 a K_0 e^{\rho}+1) \left(a K_0e^{\rho}+1\right)}\times \nonumber\\
&\frac{3 c \left(3 a K_0 e^{\rho}+1\right)+ (4 a K_0 e^{\rho}+1)^{\frac{3}{2}} \left(a K_0 e^{\rho}+3\right)}{2 \sqrt{4 a K_0 e^{\rho}+1} - a K_0 e^{\rho} \left(\sqrt{4 a K_0 e^{\rho}+1} + 3 c\right)}, \\
Y_H(\rho) =&\frac{3 a K_0 e^{\rho}}{2 \left(1+ a K_0 e^{\rho}\right)}.
\end{align}
\end{subequations}
\section{Two fluid constant density solution}\label{CD} 
As a first example of a two-fluid exact solution, we consider the simplified case in which there is no interaction and no fluxes. We assume, in addition, that the fluids have energy densities which are both constant but differ in value. In this case, the equations \eqref{tov2} can be solved directly to obtain a two-fluid generalization of Schwarzschild's interior solution.

Let us name the constant densities for fluids 1 and 2 as $\mu_1$ and $\mu_2$. In this case we have 
\begin{equation}
\mathbb{M}_i= \mu _i K_0 e^{\rho} \mathcal{K}, \quad i=1,2\,.
\end{equation}
Substituting this into \eqref{K2}, we obtain the solution:
\begin{equation} \label{KMuConstFl}
\begin{split}
\mathcal{K}=& \frac{3}{4 \mathfrak{z}^2},\\
\mathfrak{z}=& \sqrt{3 - \mu_{tot} K_0 e^{\rho}},
 \end{split}
\end{equation}
which, solving the equation for the total pressure \eqref{EqPtot} leads to
\begin{equation}
P_{tot}=-\frac{\mu_{tot} K_0 e^{\rho} (3 \mathfrak{z}+{\mathfrak c}_{tot})}{4 \mathfrak{z}^2 (\mathfrak{z}+{\mathfrak c}_{tot})} ,
\end{equation}
where $\mu_{tot} =\mu _1+ \mu _2$ and ${\mathfrak c}_{tot}$ is an integration constant. We now substitute in the first of Eqs.~\eqref{tov2} the relation $P_2=P_{tot} - P_1$ and the solution above to obtain $P_1$ and $P_2$: 
\begin{align}
P_1(\rho)=&\frac{e^{\rho} \left(4 {\mathfrak c}_{1} - 3 \mu _1 K_0 \mathfrak{z} \right)}{4 \mathfrak{z}^2 \left(\mathfrak{z}+{\mathfrak c}_{tot}\right)},\\
P_2(\rho)=&-\frac{e^{\rho} \left(4 {\mathfrak c}_{1} - 3 K_0 \mathfrak{z} \mu_1 + 
\mu_{tot} K_0 (3 \mathfrak{z} + {\mathfrak c}_{tot})\right)}{4 \mathfrak{z}^2 ( \mathfrak{z} +{\mathfrak c}_{tot})},
\end{align}
where ${\mathfrak c}_{1}$ and ${\mathfrak c}_{2}$ are constants such that ${\mathfrak c}_{tot}={\mathfrak c}_{1}+{\mathfrak c}_{2}$. Mapping $\rho$ to the area radius and utilizing
\begin{equation}
\phi = -\frac{2 \mathfrak{z}}{\sqrt{3 K_0 e^{\rho}}},
\end{equation}
we can calculate the pressures $p_1$ and $p_2$ as
\begin{subequations}\label{cdps}
\begin{align}
p_{1}(r) &= \frac{4 {\mathfrak c}_{1} -3 \mu_1 {r_0}^2 z}{3 {r_0}^2 \left(z + {\mathfrak c}_{tot} \right)}, \\
p_{2}(r) &= -\frac{4 {\mathfrak c}_{1} + {r_0}^2 (3 z \mu_{2} + {\mathfrak c}_{tot} \mu_{tot})}{3{r_0}^2 \left(z + {\mathfrak c}_{tot} \right)},
\end{align}
\end{subequations}
where $r_0=K_0^{1/2}$ and 
\begin{equation} \label{cdsub}
z= \sqrt{3-r^2 \mu_{tot}}.
\end{equation}
The variable $Y$ can be found using \eqref{c2}. Now, breaking covariance and choosing a metric of the form \eqref{metrick}, the expressions \eqref{vars} give
\begin{subequations}\label{cdmf}
\begin{align}
k_1 &= a_0 \left[{\mathfrak c}_{1}+{\mathfrak c}_{2}+z\right]^2, \\
k_2 &= \frac{3}{z^2},\\
k_3 &= r^2,
\end{align}
\end{subequations}
where $a_0$ is a constant and we have set $r_0=1$ without loss of generality. We will use this last convention when writing the area radius solutions in all of the following sections so that the size of the equations is reduced. 

Notice that the coefficients of the metric \eqref{cdmf}, as expected, are the same as the single fluid solution presented in the previous section. As a consequence, the central pressure will also have the same structure as the single fluid solution, and the maximum possible mass $M_{tot}$ of this object will be the same, i.e.
\begin{equation}
M_{tot}=M_1+M_2=\frac{4}{9}R.
\end{equation}
The difference is that the maximum mass can be achieved with different combinations of the two fluids.

\begin{figure}
\centering
\includegraphics[scale=0.65]{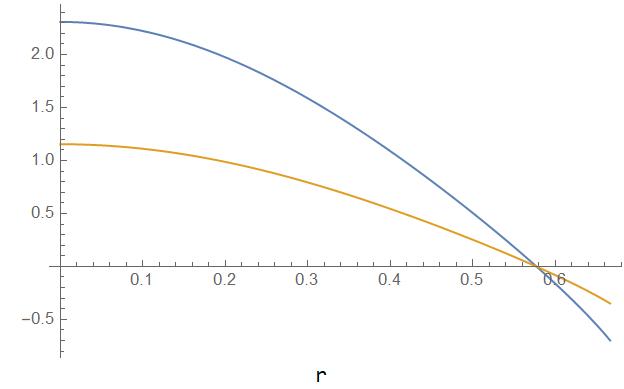}
\caption{Pressure for fluid 1 (blue) and fluid 2 (orange) vs $r$ for the double constant density solution found in Section \ref{CD} with $\mu_1 = 4$ and $\mu_2 = 2$, and parameter values ${\mathfrak c}_{1} = 3$ and ${\mathfrak c}_{2} = -6$.}
\label{fig1}
\end{figure}

In Fig. \ref{fig1} we give an example of the behaviour of the pressures $p_1$ and $p_2$ in a specific case that is compatible with the requirements given in Sec. \ref{CON}.
\section{Solution reconstruction}\label{SR}
In this section, we will expand the technique proposed in \cite{sante1,sante2} to deduce new two-fluid solutions starting from a given metric. We begin with rearranging (\ref{DE2}) as
\begin{equation}
X = - \frac{1}{2} - \frac{\mathcal{K}_{, \rho}}{2 \mathcal{K}}. \label{X}
\end{equation}
Using (\ref{c1}) and (\ref{c2}) we obtain
\begin{equation}
\mathbb{M}_{tot}=\mathbb{M}_1 + \mathbb{M}_2 = \frac{\mathcal{K}_{, \rho}}{2 \mathcal{K}} + \mathcal{K} - \frac{1}{4}. \label{92}
\end{equation}
Equation (\ref{DE1}) together with (\ref{92}) gives
\begin{eqnarray}
{P}_{tot}=P_1 + P_2 &=& \frac{1}{3} \left[2 Y_{, \rho} + 2 Y^2 + Y \right] - \frac{1}{3} \mathcal{K} \nonumber\\
&&- \left(\frac{2 Y + 1}{6} \right)\frac{\mathcal{K}_{, \rho}}{\mathcal{K}}+ \frac{1}{12}. \label{93}
\end{eqnarray}
Given a metric, and therefore the functions $Y$ and $\mathcal{K}$ which satisfy the relation \cite{sante1,sante2}
\begin{equation}\label{ConstrRec}
(2 Y+1) \mathcal{K}_{,\rho } - 4 \mathcal{K}^2-\mathcal{K} \left[4
 Y_{,\rho } + 4 (Y - 1) Y - 1\right]=0,
\end{equation}
we can obtain the sum of the pressure and energy density variables of the two fluids.

One could then be tempted to choose the behaviour for the energy density and pressure of one of the two fluids and deduce the other. This, however, would be a mistake, as it would ignore the conservation laws associated to the single fluids. Such an additional constraint can be introduced considering also Eq. \eqref{DE3}:
\begin{equation}\label{RecP1p}
P_{1,\rho}=-P_1 (2 X+Y)-\mathbb{M}_1 Y,
\end{equation}
which is a first order differential equation. Solving for $\mathbb{M}_1$ we obtain
\begin{equation}\label{RecP1mu}
\mathbb{M}_1=-\frac{P_1 (Y+2 X) +P_{1,\rho}}{Y},
\end{equation}
which can be used to derive $\mathbb{M}_1$ once $P_1$ is given. In fact, remembering the formulae in \eqref{vars}, Eq. \eqref{RecP1p} can be solved in general in terms of the metric coefficients and the energy density to give
\begin{equation}
P_1=\frac{k_2^2}{\sqrt{k_1}} \left({\mathfrak c}_{1} - \int \frac{\mu _1 k_{1,\rho }}{2 \sqrt{k_1} k_2}d\rho\right). 
\end{equation}
With this result, one can assign the energy density and derive the pressure. Clearly, because \eqref{RecP1p} is an ordinary differential equation, the two approaches above are completely equivalent. 

Hence, once we choose a geometry, we can then choose the behavior for the energy density or pressure of one of the two fluids and deduce the other quantities via the elementary relations
\begin{subequations}
\begin{align}
\mathbb{M}_{2} &= \mathbb{M}_{tot} - \mathbb{M}_{1}, \\
P_{2} &= P_{tot} - P_{1}.
\end{align}
\end{subequations}
At this point, one can obtain the expressions for the energy densities and the pressures, taking into account that the factor $\phi^2$ is the one associated with the underlying geometry represented by $\mathcal{K}$ and $Y$ in the formulae above.

In the following, we will construct models of a relativistic star with two fluids using the Tolman IV and Heintzmann geometry and employing $P$ and $\mathbb{M}$ of the interior Schwarzschild, Tolman IV and Heintzmann solutions\footnote{It should be noted here that, because of the different forms of $\phi$, the relation between $P$ and $\mathbb{M}$ with $p$ and $\mu$ is not the usual one. For example, in the Tolman IV-Heintzmann object, $P_T$ and $\mathbb{M}_T$ will not correspond to $p_T$ and $\mu_T$. These last quantities are in fact connected to $p_1$ and $\mu_1$ by
\begin{equation}
p_1=p_T \frac{\phi^2_H}{\phi^2_T}, \quad \mu_1=\mu_T \frac{\phi^2_H}{\phi^2_T}.
\end{equation} 
In this sense, the name ``Tolman IV-Heintzmann object'' will indicate the origin of the solution, rather than its actual composition.}. These choices are particularly convenient as the constraint \eqref{ConstrRec} is already satisfied. However, the algorithm is general and one can use any other known physical solution to obtain two-fluid solutions. Naturally, these new solutions should be tested against the requirements given in Sec. \ref{CON} to verify their compatibility with physical objects. 

\subsection{Tolman IV-Constant density relativistic object}

\begin{figure}
\centering
\includegraphics[scale=0.65]{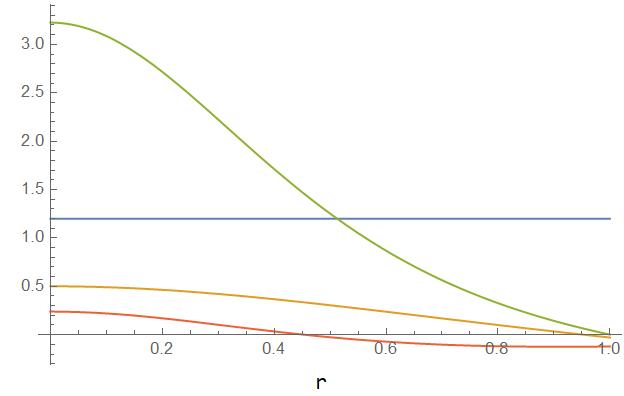}
\caption{The energy density (blue) and pressure (orange) for fluid 1, and the energy density (green) and pressure (red) for fluid 2 vs $r$ for the Tolman IV-constant density object. We used the parameter values $A = 0.95 , R = 1.65, \mu_{1} = 1.2,{\mathfrak c}_1= -1.1$, which satisfy the energy and stability conditions \eqref{energycond}, \eqref{stabcond} and \eqref{pcond}, in order to obtain the clearest representation. The pressure for fluid 2 becomes negative close to $r=0.455$, as indicated by the dashed vertical line.}
\label{fig2}
\end{figure}

\begin{figure}
\centering
\includegraphics[scale=0.65]{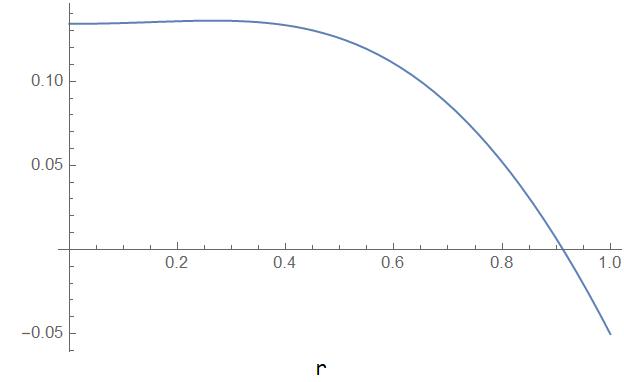}
\caption{The plot of the square of the sound speed ($\frac{\partial p}{\partial \mu}$) for the Tolman IV-constant density object. We used the parameters values $A = 0.95 , R = 1.65, \mu_{1} = 1.2,{\mathfrak c}_1= -1.1$ as in Fig. \ref{fig2}}
\label{fig2a}
\end{figure}

We start with the Tolman IV geometry (\ref{tol4}) and choose one of the fluids to be the constant density fluid in Sec. \ref{CD}. In this way the energy density and the pressure of the two fluids will be given by
\begin{subequations}\label{combined11}
\begin{align}
\mu_{2} &= \frac{3 A^4+2 r^2 \left(3 r^2+R^2\right) + A^2 \left(7 r^2+3 R^2\right)}{R^2 \left(A^2+2 r^2\right)^2}- \mu_1, \\
p_1&= -\mu_1-\frac{4 \mathfrak{c}_1}{R^2 \sqrt{A^2+r^2}},\\
p_{2} & = \frac{-A^2-3 r^2+R^2}{R^2 \left(A^2+2 r^2\right)}+\frac{4 \mathfrak{c}_1}{R^2 \sqrt{A^2+r^2}}+\mu_1,
\end{align}
\end{subequations}
where $\mu_1$ is the constant density of the first fluid, $\mathfrak{c}_1$ is a constant of integration and we have used \eqref{tol4mv}.
The properties of the underlying geometry, the energy density and pressure of the first fluid guarantee that $\mu_{2}$ and $p_{2}$ do not have singularities. Naturally we should guarantee that the conditions \eqref{energycond}, \eqref{stabcond} and \eqref{pcond} are all satisfied. Given the number of parameters it is not practical to give these conditions analytically. Hence we limit ourselves to show graphically that at least one of these combinations exists. Figure \ref{fig2} shows the pressure and energy density profiles for both fluids and Fig. \ref{fig2a} shows the square of the sound speed, $\frac{\partial p}{\partial \mu}$.

Notice that unlike in Fig. \ref{fig1}, the pressures do not approach zero at the same value of the radial coordinate. Indeed, we could find no value of the parameters for which $p_1$ and $p_2$ are zero at the same $r$. However, this does not imply that the solution we found should be discarded: for the interval in which $p_1$ and $p_2$ are both positive, the combination of \eqref{tol4} with \eqref{combined11} constitutes an acceptable two fluid solution. It is clear that, in order to obtain a complete model of a compact object, such a solution should be joined, using e.g. Israel's prescriptions \cite{Israel:1966rt,Barrabes:1991ng}, to a shell or another solution which satisfies the conditions (\ref{energycond}-\ref{pcond}), and whose pressure(s) approach zero at some value of the radial coordinate $r$.

In other words, we obtain naturally a shelled structure for this object: an internal shell in which two fluids are present and an external one which can be matched smoothly to the Schwarzschild solution. To model the external shell we can use any geometry or number of fluids. An easy setting for the external shell could be a single fluid Tolman IV solution which has pressure $p_1$ and energy density $\mu_1+\mu_2$ at the junction point (where $p_2$ vanishes). In the following, we will often find ourselves in the same situation. For the sake of brevity, we will not give a description of the single fluid external shells for each instance.

\subsection{Heintzmann-Constant density relativistic object}
\begin{figure}
\centering
\includegraphics[scale=0.65]{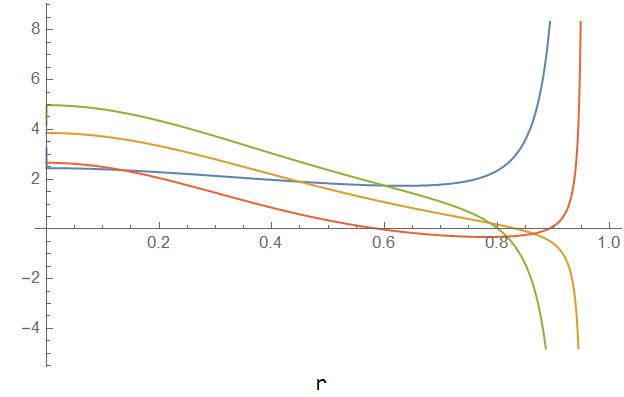}
\caption{The energy density of Fluid 1 (blue), and Fluid 2 (green), as well as the pressure for Fluid 1 (orange), and for Fluid 2 (red), vs $r$ for the Heintzmann-constant density object. We used the parameter values $a = 1.5, c = 0.1, \mu_{1} = 3.3, {\mathfrak c}_1 = -2.5$, which satisfy the energy and stability conditions \eqref{energycond}, \eqref{stabcond} and \eqref{pcond}, in order to obtain the clearest representation. The pressure of Fluid 2 becomes negative close to $r=0.6$.}
\label{fig3}
\end{figure}

\begin{figure}
\centering
\includegraphics[scale=0.65]{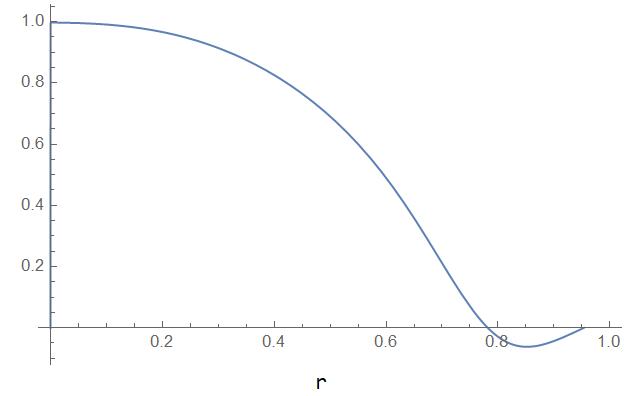}
\caption{The plot of the square of the sound speed ($\frac{\partial p}{\partial \mu}$) for the Heintzmann-constant density object. We used the parameters values $a = 1.5, c = 0.1, \mu_{1} = 3.3, {\mathfrak c}_1 = -2.5$ as in Fig. \ref{fig3}}
\label{fig3a}
\end{figure}

We start with the Heintzman IIa geometry (\ref{hein2a}) and choose one of the fluids to be the constant density fluid in Sec. \ref{CD}. In this way the energy density and the pressure of the two fluids will be given by 
\begin{subequations}\label{combined33}
\begin{align}
\mu_{1} &= \frac{\mu_1 z_2 z_3}{z_1^4 ({\mathfrak c}_1+z_1)^2} \left(2 {\mathfrak c}_1 \mu_1 r^2 ({\mathfrak c}_1+3 z_1) \right.\nonumber \\
&\left. + z_1^2 (z_4 ({\mathfrak c}_1+z_1) ({\mathfrak c}_1+3 z_1)+2 {\mathfrak c}_1 ({\mathfrak c}_1+4 z_1)+18)\right),\\
\mu_{2} &=-z_3 z_5 + \frac{\mu_1 z_2 z_3}{z_1^4 ({\mathfrak c}_1+z_1)^2} \left(2 {\mathfrak c}_1 \mu_1 r^2 z_1 (2 z_4+1) \right.\nonumber \\
&\left. -z_1 z_4 \left({\mathfrak c}_1^2 z_1+12 {\mathfrak c}_1+3 z_1^3\right)-6 ({\mathfrak c}_1+z_1) ({\mathfrak c}_1+3 z_1)\right),\\
p_1 &= -\frac{\mu_1 z_3 ({\mathfrak c}_1+3 z_1)}{z_1^2 ({\mathfrak c}_1+z_1)},\\
p_{2} & = z_3 \left(\frac{\mu_1 ({\mathfrak c}_1+3 z_1)}{z_1^2 ({\mathfrak c}_1+z_1)}+z_6 \right),
\end{align}
\end{subequations}
where we have written
\begin{eqnarray} \label{zz}
z_0 &=& \sqrt{4 a r^2+1}, \nonumber\\
z_1 &=& \sqrt{3-\mu_1 r^2}, \nonumber\\
z_2 &=& \frac{1 + a r^2}{3 a r^2}, \nonumber\\
z_3 &=& 1-\frac{3\left(z_0+ c\right)} {2 z_2z_0}, \nonumber \\
z_4 &=& \frac{a r^2 \left\{a r^2 \left[4 a z_0 r^2+9 (z_0-c)\right]+18 z_0\right\}+4 z_0}{z_0\left(a r^2+1\right) \left[a r^2 \left(3 c z_0+4 a r^2-7\right)-2\right]}, \nonumber \\
z_5 &=& \frac{3 a \left[c \left(9 a r^2+3\right)+\left(a r^2+3\right) z_0^3\right]}{z_0^2 \left(a r^2+1\right) \left[a r^2 \left(z_0+3 c\right)-2 z_0\right]}, \nonumber \\
z_6 &=& \frac{3 a \left[7 a c r^2+3 \left(a r^2-1\right) z_0+c\right]}{\left(a r^2+1\right) \left[a r^2 \left(z_0+3 c\right)-2 z_0\right]}, \nonumber \\
\end{eqnarray}
for compactness, $\mu_1$ is the energy density associated with the constant density solution, ${\mathfrak c}_1$ is a constant of integration, and we have used \eqref{hein2a}. Figure \ref{fig3} shows the pressure and energy density profiles of both fluids, and Fig. \ref{fig3a} shows $\frac{\partial p}{\partial \mu}$.

\subsection{Heintzmann-Tolman relativistic object}
\begin{figure}
\centering
\includegraphics[scale=0.65]{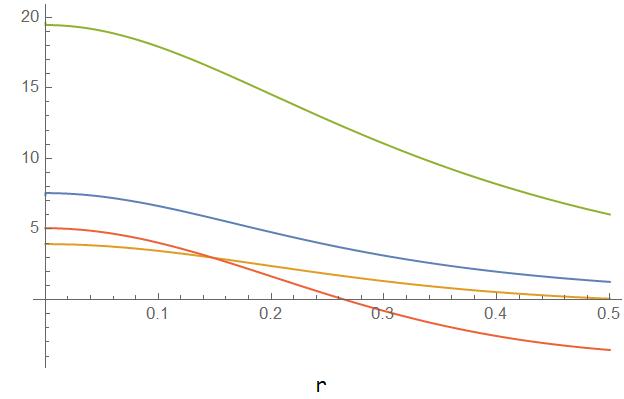}
\caption{The energy density of Fluid 1 (blue), and Fluid 2 (green), as well as the pressure for Fluid 1 (orange), and for Fluid 2 (red), vs $r$ for the Heintzmann-Tolman object. We used the parameter values $A = 0.5, R = 4, a = 3, c = 1$, which satisfy the energy and stability conditions \eqref{energycond}, \eqref{stabcond} and \eqref{pcond}, in order to obtain the clearest representation. The pressure of Fluid 2 becomes negative close to $r=0.25$.}
\label{fig4}
\end{figure}

\begin{figure}
\centering
\includegraphics[scale=0.65]{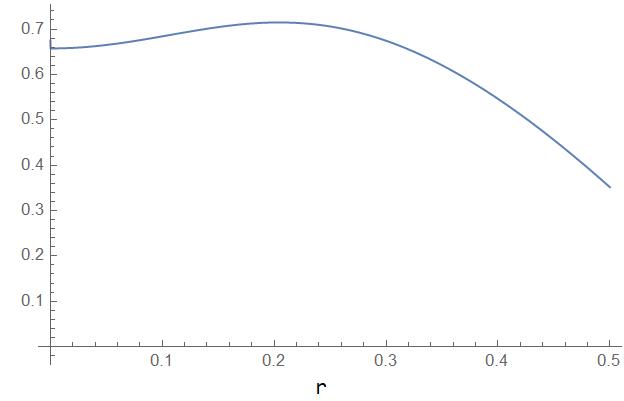}
\caption{The plot of the square of the sound speed ($\frac{\partial p}{\partial \mu}$) for the Heintzmann-Tolman object. We used the parameter values $A = 0.5, R = 4, a = 3, c = 1$ as in Fig. \ref{fig4}.}
\label{fig4a}
\end{figure}

We start from the Heintzmann IIa geometry (\ref{tol4}) and choose one of the fluids to be the Tolman IV fluid (\ref{hein2a}). In this way the energy density and the pressure of the two fluids will be given by 
\begin{subequations}\label{combined22}
\begin{align}
\mu_{1} &= \frac{z_2 z_3}{z_7^2}\left[-R^4 \left(A^2 (z_4+2)+z_4 r^2\right) -4 A^2 r^4(z_4+1)\right. \nonumber\\
& \left. + R^2 \left(A^2 r^2 (5 z_4+12) +A^4 (z_4+2)+4 r^4 (z_4+1)\right)\right. \nonumber\\
& \left. -A^4 r^2 z_4-3 r^6 z_4 \right], \\
\mu_{2} &= z_3 \Bigg(-z_5+\frac{z_2}{z_7^2 (r+R)^2} \times \Bigg[3 z_7 r^2 (z_4+4)\Bigg. \Bigg.\nonumber\\
& \Bigg. \Bigg. - R^2 z_7(z_4+2) + A^2 \left(z_7(z_4+2) +4 r^2 R^2-10 r^4\right)\Bigg. \Bigg.\nonumber\\
& \Bigg. \Bigg. -2 A^4 r^2 -2 r^2 \left(-5 r^2 R^2+6 r^4+R^4\right)\Bigg] \Bigg), \\
p_1 &=\frac{z_8 z_3}{z_7}, \\
p_{2} &= \frac{z_3 (z_6 z_7-z_8)}{z_7},
\end{align}
\end{subequations}
where we have used the $z_i$ given in \eqref{zz} and written
\begin{equation}
\begin{split}
z_7 &= \left(A^2+r^2\right) \left(r^2-R^2\right), \\
z_8 &= A^2+3 r^2-R^2,
\end{split}
\end{equation}
for compactness and we have used \eqref{tol4mv} and \eqref{hein2amv}. Figure \ref{fig4} shows the pressure and energy density profiles of the two fluids, and Fig. \ref{fig4a} shows the square of the sound speed, $(\frac{\partial p}{\partial \mu})$.

\section{A generating theorem}\label{GT}
In their 2005 paper, Boonserm {\it et al.} \cite{boon1, boon2} developed transformation theorems able to map one perfect fluid sphere into another called {\it generating theorems}. The work in \cite{boon1, boon2} focuses on the spacetime geometry, starting with a known perfect fluid sphere and applying the theorem in order to obtain a new class of solutions to the Einstein field equations. Certain metrics, known as seed metrics, are transformed to generate new metrics. 

In \cite{sante1,sante2} it was shown that in the context of the covariant version of the TOV equations, the generating theorems assume the simple forms of linear solution deformation. In this section, we will show that one of such theorems can also be used to obtain two-fluid solutions from single fluid ones. In the following, we will focus on the simplest cases in which the solution can be obtained analytically. In doing so, we will assume isotropy, no fluxes and no interaction. As seen in the previous section, and as it happens for the single fluid generating theorems, it is not obvious that the new solution will be compatible with the conditions \eqref{energycond}, \eqref{stabcond} and \eqref{pcond}. These constraints must be imposed {\it a posteriori} to evaluate the physical relevance of the solution \footnote{It is worth adding that relaxing the requirement of deducing an analytical form extends enormously the number of achievable solutions. We will not undertake such analysis here. The reader is reminded, however, that such course of action implies the additional problem to find a justification for the chosen values of the parameters.}.

The TOV equations in the case $Q_1=0$ reads
\begin{equation}\label{tovSimple}
\begin{split}
P_{,\rho} =& -P^2 +P \left[\mathbb{M}+ 1 -3 \left(\mathcal{K} - \frac{1}{4} \right) \right] \\
&- \left(\mathcal{K} - \frac{1}{4}\right) \mathbb{M}, \\
\mathcal{K}_{,\rho} =& - 2 \mathcal{K}\left(\mathcal{K}- \frac{1}{4} - \mathbb{M} \right). 
\end{split}
\end{equation}
Given a solution of (\ref{tovSimple}) represented by $P_0, \mathbb{M}_0, \mathcal{K}_0,$ and $Y_0$ we perform the linear deformation
\begin{equation}\label{def1}
\begin{split}
P_0 &\rightarrow P_0 + \tilde{P}, \quad Y_0 \rightarrow Y_0 + \tilde{Y},\\
\mathbb{M}_0 &\rightarrow \mathbb{M}_0, \quad\quad~~ \mathcal{K}_0\rightarrow \mathcal{K}_0
\end{split}
\end{equation}
corresponding to Theorem 2 of \cite{boon1}. We can use this theorem to obtain a two fluid solution from a single fluid one. 

In fact, the transformation \eqref{def1} can be interpreted as a change from a setting in which we have a {\it single} fluid with pressure $p_0= P_0\phi^2$ to a new setting in which we have {\it two fluids} of pressures $p_0= P_0\phi^2$ and $\tilde{p}= \tilde{P}\phi^2$ where $\tilde{P}$ is given by the equation
\begin{equation}\label{def1tol4}
\tilde{P}_{,\rho} + \tilde{P}^2 + \tilde{P} \left(3 \mathcal{K}_0 - \mathbb{M}_0 + 2 P_0 - \frac{7}{4} \right) = 0, 
\end{equation}
derived by combining \eqref{DE1}, \eqref{DE2} and the constraints \eqref{constraints} with the transformation \eqref{def1}.
 
We then need to determine the energy density of the two fluids. This can be accomplished using the conservation law
\begin{equation}\label{constrdef}
\tilde{P}_{,\rho}=-\tilde{P} (2 X+Y)-\tilde{\mathbb{M}} Y,
\end{equation}
that has to hold for the fluid with pressure $\tilde{p}$. Combining \eqref{DE1}, \eqref{DE2} and the constraints \eqref{constraints} with the transformation \eqref{def1} and \eqref{constrdef}, we obtain $\tilde{Y} = \tilde{P}$ and
\begin{equation}
\tilde{\mathbb{M}}=\frac{\tilde{P} \left(1-4 \mathcal{K}_0+4 \mathbb{M}_0+4 Y_0\right)}{4
 \left(\tilde{P} + Y_0\right)},\label{GenMat1}
\end{equation}
This last equation allows to obtain the energy density corresponding to $\tilde{p}$ as $\tilde{\mu}= \tilde{\mathbb{M}}\phi^2$. Instead, the energy density of the remaining fluid can be obtained keeping in mind that the total energy density should be $\mu_0$. From \eqref{def1} we can then write
\begin{equation}\label{defbarM0}
\mathbb{M}_0=\bar{\mathbb{M}}_0+\tilde{\mathbb{M}}\;,
\end{equation}
where $\bar{\mathbb{M}}_0$ is the energy density variable associated to the fluid with pressure $p_0$ in the two fluid picture. The energy density for this last fluid will be, then,
\begin{equation}\label{defbarmu0}
\bar{\mu}_0=\mu_0-\tilde{\mu}
\end{equation}
where $\bar{\mu}_0=\bar{\mathbb{M}}_0\phi^2$. Hence starting with a known single fluid solution characterised by $(p_{0},\mu_0)$ we obtain a two fluids solution characterised by a fluid with $(\tilde{p},\tilde{\mu})$ and a second fluid with $(p_{0},\bar{\mu}_0)$. 

The transformation \eqref{def1} induces a transformation of the metric coefficients given by
\begin{subequations}\label{def1tol4mc}
\begin{align}
k_1 &\rightarrow k_{1} \exp \left(\int \tilde{Y} d\rho \right)\,,\\
k_2 &\rightarrow k_{2}\,, \\
k_3 &\rightarrow k_{3}\,.
\end{align}
\end{subequations}
where we have used the form \eqref{metrick}. These relations allow to complete the description of the new two fluid solution.

\begin{figure} 
\centering
\includegraphics[scale=0.65]{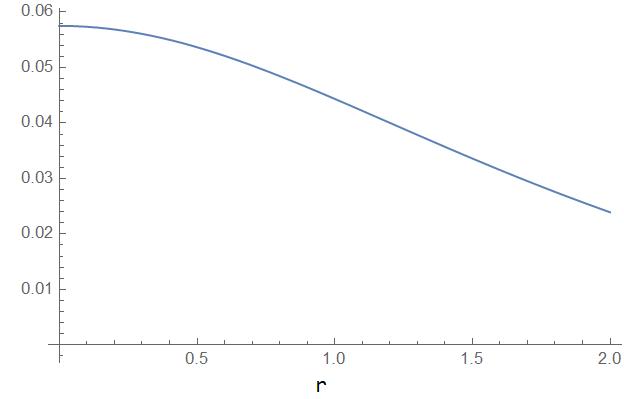}
\caption{Pressure $\tilde{p}$ vs $r$ for the solution generated in Section \ref{GT} with constant values $A = -2.55, R = -3.6, {\mathfrak c}_1 = -3.6$.}
\label{fig5}
\end{figure}

\begin{figure} 
\centering
\includegraphics[scale=0.65]{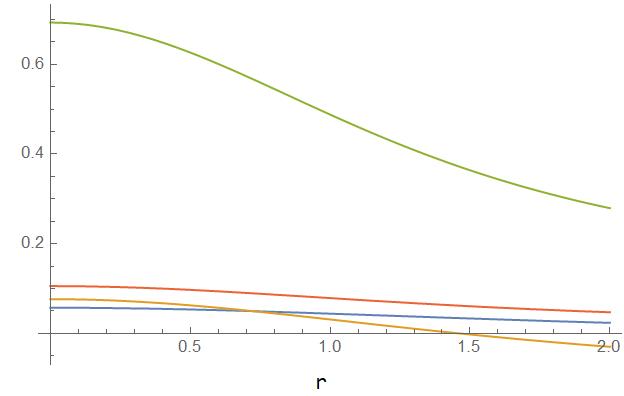}
\caption{The pressure $p_0$ (green), deformation pressure $\tilde{p}$ (blue), energy density $\bar{\mu}_0$ (red), and deformation energy density $\tilde{\mu}$ (orange) vs $r$ for the solution generated in Section \ref{GT} with constant values $A = -2.55, R = -3.6, {\mathfrak c}_1 = -3.6$.}
\label{fig6}
\end{figure}

\begin{figure} 
\centering
\includegraphics[scale=0.65]{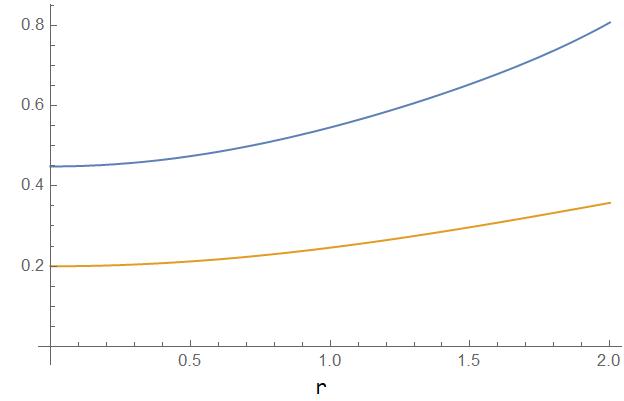}
\caption{The square of the sound speed ($\frac{\partial p}{\partial \mu}$) for the fluid with pressure $p_0$ (orange), and with $\tilde{p}$ (blue) vs $r$ for the solution generated in Section \ref{GT} with constant values $A = -2.55, R = -3.6, {\mathfrak c}_1 = -3.6$.}
\label{fig6a}
\end{figure}

\begin{figure}
\centering
\includegraphics[scale=0.65]{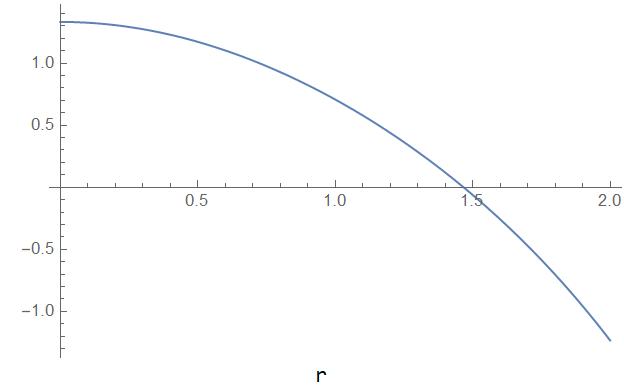}
\caption{$p_0 / \tilde{p}$ vs $r$ for the solution generated in Section \ref{GT}, with constant values $A = -2.55, R = -3.6, {\mathfrak c}_1 = -3.6$.}
\label{fig7}
\end{figure}

\begin{figure} 
\centering
\includegraphics[scale=0.65]{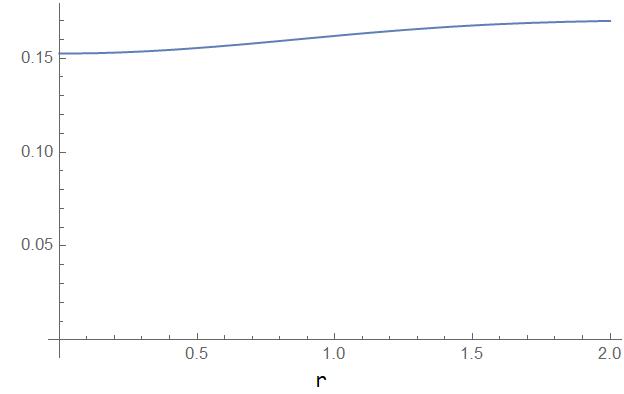}
\caption{$\tilde{\mu} / \bar{\mu}_0 $ vs $r$ for the solution generated in Section \ref{GT} with constant values $A = -2.55, R = -3.6, {\mathfrak c}_1 = -3.6$.}
\label{fig8}
\end{figure}

Let us now suppose that the starting solution is the Tolman IV metric given in \eqref{tol4nmv}. The formulas above lead to a new solution which is sourced by a first fluid whose pressure is given by 
\begin{align}
\tilde{p} (r) =& \frac{4 z_1 \sqrt{A^2+r^2} \left(A^2+R^2\right)}{z_4 R^2} \times \nonumber\\
& \left\{\left[2 z_2 z_3 + {\mathfrak c}_1 \left(A^2 + R^2\right)\right]\left(A^2+r^2\right)\right.\nonumber\\
&\left. +2 z_1 z_4 \sqrt{A^2+r^2}\right\}^{-1},
\end{align}
where
\begin{subequations}\label{alphas}
\begin{align}
z_1 =& \sqrt{R^2-r^2}, \\
z_2 =& E\left(\sin ^{-1}\left(\sqrt{\frac{R^2-r^2}{A^2+R^2}}\right),\frac{A^2}{A^2+2 R^2}+1\right),\nonumber\\
& -F\left(\sin ^{-1}\left(\sqrt{\frac{R^2-r^2}{A^2+R^2}}\right),\frac{A^2}{A^2+2 R^2}+1\right),\\
z_3 =& \sqrt{A^2+2 R^2},\\
z_4 =& \sqrt{A^2+2 r^2}
\end{align}
\end{subequations} 
and ${\mathfrak c}_1$ is a constant, $E$ is the complete elliptical integral and $F$ is the elliptical integral of the first kind. The corresponding energy density is given by
 equation \eqref{GenMat1} and reads:
\begin{align}
\tilde{\mu}(r) =& \frac{4 z_3^2 \sqrt{A^2+r^2} \left(A^2+R^2\right)}{R^2 z_4^3} \times\nonumber\\
& \Big\{\left(2 z_4 \sqrt{A^2+r^2} + {\mathfrak c}_1 z_1 \left(A^2+R^2\right) \right. \nonumber \\
& \left. -2 z_1 z_2 z_3 \right)\Big\}^{-1},
\end{align}
where the quantities $z_i$ are the same as \eqref{alphas}. Instead the second fluid will have pressure $p_0$ given by \eqref{tol4p} i.e.
\begin{equation}
p_0=p_T= \frac{R^2-A^2-3 r^2}{R^2 \left(A^2+2 r^2\right)}
\end{equation}
and energy density given by \eqref{defbarmu0} i.e.
\begin{align}
\bar{\mu}_0=& \frac{R^2 \left(3 A^2+2 r^2\right)+7 A^2 r^2+3 A^4+6 r^4}{R^2 \left(A^2+2 r^2\right)^2} \\
& - \frac{4 z_3^2 \sqrt{A^2+r^2} \left(A^2+R^2\right)}{R^2 z_4^3} \times\nonumber\\
& \Big\{\left(2 z_4 \sqrt{A^2+r^2} + {\mathfrak c}_1 z_1 \left(A^2+R^2\right) \right. \nonumber \\
& \left. -2 z_1 z_2 z_3 \right)\Big\}^{-1}.
\end{align}
Figure \ref{fig5} shows a plot of the pressure, $\tilde{p}$. The four quantities $\mu_0$, $\tilde{\mu}$, $p_0$, $\tilde{p}$ are graphically represented in Fig. \ref{fig6}, and the equations of state are shown graphically in Fig. \ref{fig6a}, for specific values of the parameters . We also give a plot of the the ratio of the two pressures $p_0 / \tilde{p}$, shown in Fig. \ref{fig7} and the two energy densities in Fig. \ref{fig8}.

\begin{figure} 
\centering
\includegraphics[scale=0.65]{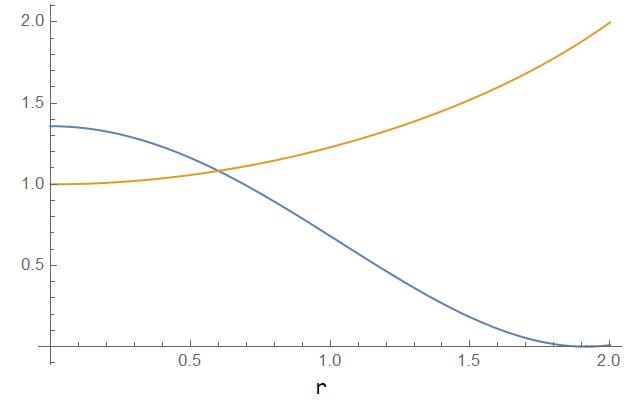}
\caption{The metric coefficients $k_{1}$ (blue) and $k_{2}$ (orange) vs $r$ for the solution generated in Section \ref{GT}. We use the constant values $A = -2.55, R = -3.6, {\mathfrak c}_1 = -3.6$ and ${\mathfrak c}_2 = -1.45$.}
\label{fig9}
\end{figure}

Finally, using \eqref{def1tol4mc}, the full expression of the metric coefficients is found to be
\begin{align}
k_1=&\Bigg[\frac{2 {\mathfrak c}_1 }{A^2+R^2}\left(\frac{z_1 z_4}{\sqrt{A^2+r^2}}-z_2 z_3 \right)\nonumber\\
 &+\sqrt{A^2+r^2}+ {\mathfrak c}_2 \Bigg]^2,\\
k_2=& \frac{R^2{z_4}^2}{{z_1}^2 \left(A^2 + r^2 \right)}, \\ 
k_3=&r^2,
\end{align}
where the $z_i$ are the same as \eqref{alphas}. We show a plot of these metric coefficients in Fig. \ref{fig9}.

\section{Introducing the fluxes}\label{SecQ}
Looking at equation \eqref{Q2} it is clear that a two fluid static solution can be of two types. The first, in which the fluxes are identically zero and a second, in which the flux of one fluid counteracts exactly the flux of the other. In this section we will consider this last class of solutions. The analysis of the behaviour of the fluxes in these solutions is relatively straightforward. Equations \eqref{AphiCoord} allow us to write equation \eqref{Q1} as 
\begin{equation}\label{bbQ_Sol}
\hat{Q}_1= -Q_1\left(\frac{\hat{k}_3}{k_3}+ \frac{\hat{k}_1}{k_1}\right),
\end{equation}
where we have set the interaction terms to zero. The above equation can be integrated by separation of variables to give
\begin{equation}\label{SolQ}
{Q}_1=\frac{\bar{Q}_1}{k_1k_3}=- Q_2,
\end{equation}
where $\bar{Q}_1$ is an integration constant. The same expression holds in terms of the parameter $\rho$ and the area radius $r$ as $k_1$ and $k_3$ are scalars for changes in these parameters. Notice that the behavior of the flux depends only on the spacetime geometry, and therefore it is independent of the properties of matter. 

An important aspect of equation \eqref{SolQ} is that, as the flux is inversely proportional to the Gaussian curvature, we expect that in the center, where $k_3=0$ (the metric \eqref{metrick} is singular), the flux will diverge. Such divergence is present also without spherical symmetry and even if we relax the assumption of staticity. In fact, in the case of a completely general LRS class II spacetime of the type \eqref{metrick4} the 1+1+2 equations give \cite{betschart}
\begin{equation}
Q= \frac{\hat{\dot{k}}_3}{k_3}-\frac{\dot{k}_3}{k_3}\frac{\hat{k}_2}{k_2}-\frac{\dot{k}_3}{k_3}\frac{\hat{k}_3}{k_3},
\end{equation}
which might also diverge when $k_3=0$. The origin of this divergence is fundamentally related to the pathological behaviour of the angular coordinates in $r=0$ and therefore we might expect that using a different chart the divergence of $Q$ would disappear, not unlike the case of the divergence of the Schwarzschild horizon.

\begin{figure} 
\centering
\includegraphics[scale=0.65]{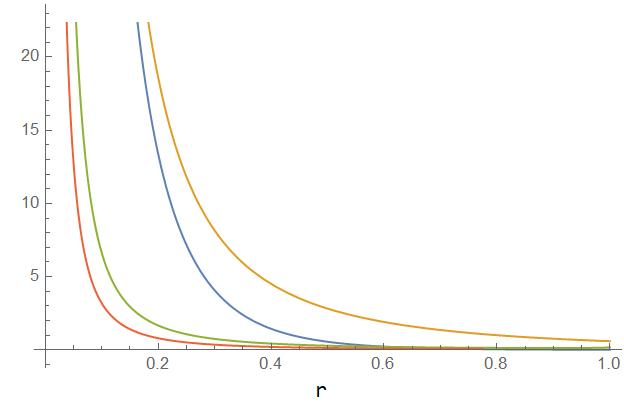}
\caption{The flux for: Interior Schwarzschild (red), Tolman IV (orange), Heint IIa (blue) and the solution found with the generating theorem (green) vs $r$. We use the constant values $\bar{Q}_1 = 3, a = 3, b = 2, A = 2, B = 2, R = 1,\mu_1 = 2, \mu_2 = 3, {\mathfrak c}_1 = 3, {\mathfrak c}_2 = 4, {\mathfrak c}_3 = 5, {\mathfrak c}_4 = 3$.}
\label{fig10}
\end{figure}

As an application we will now give the form of the fluxes for the solutions that we have found in the previous sections. For the two-fluid constant density interior Schwarzschild solution \eqref{cdmf} of Section \ref{CD} we have
\begin{equation}
{Q}_1=\frac{ \bar{Q}_1}{r^2 \left[{\mathfrak c}_{1}+{\mathfrak c}_{2}+z\right]^2}.
\end{equation}
For the solution based on the Tolman geometry in Section \ref{SR} we have
\begin{equation}
{Q}_1=\frac{A^2 \bar{Q}_1}{r^2 B^2 \left(A^2+ r^2\right)},
\end{equation}
whereas for the solution based on the Heintzmann geometry in Section \ref{SR} we have 
\begin{equation}
{Q}_1=\frac{ \bar{Q}_1}{r^2 b^2 \left(1+ a r^2\right)^3}.
\end{equation}
Finally, for the solution obtained by the generating theorem in Section \ref{GT}, we have
\begin{align}
{Q}_1 = & \frac{\bar{Q}_1 \left(A^2+R^2\right)^2}{r^2} \Bigg[-2 {\mathfrak c}_1 z_2 z_3 \sqrt{A^2+r^2} \nonumber \Bigg. \\
&+ \Bigg. {\mathfrak c}_2 \left(A^2+R^2\right)+2 {\mathfrak c}_1 z_1 z_4 \Bigg]^{-2},
\end{align}
where the $z_i$ are given by \eqref{alphas}. The fluxes are graphically represented in Fig. \ref{fig10}.
\section{Discussion and conclusion}\label{CR}
In this paper, we have presented a complete set of equations able to describe the interior solution of relativistic objects with more than one fluid source, including possible anisotropic stress, energy flux, and interaction terms. These equations have been written by means of the 1+1+2 covariant formalism, which allows a relatively straightforward treatment of the many features of these kinds of systems.  As it is done in GR, the 1+1+2 equations can be combined to obtain the covariant equivalent of the TOV equations.

The properties of the TOV equations, however, can be more clearly appreciated when they are of recast in a form that either contains dimensionless variables or it is written in terms of quantities which are invariant under homological transformation. Using the 1+1+2 potential we have defined variable with similar properties, which allow one to write the TOV equations as a closed system of a Riccati and a Bernoulli equations, when the equation of state of matter is included. As in the case of the single-fluid solutions, these equations can be solved exactly with several techniques, other than direct resolution. 

We have presented here some of these solutions. In particular, we have obtained the generalization of the interior Schwarzschild solution to the case of two fluids. We have also formulated the reconstruction algorithm for the two-fluid TOV equations. The structure of the reconstruction equations seems to show that there is a degeneracy as a given spacetime metric might correspond to many different combinations of fluids. The origin of such degeneracy is ultimately correlated with the equivalence principle: as all matter gravitates and does it in the same way, several multi-fluid configurations can produce the same spacetime metric. However, in the two-fluid case, the conservation laws for the individual fluids are independent equations, and therefore one must include an additional equation. Consequently, an additional constraint must be added to the reconstruction equations and this additional equation immediately resolves the degeneracy. We found that the additional equation can be seen as a differential constraint for the pressure of one of the fluids or an algebraic one for the energy density. Using the new reconstruction equation and the single fluid solutions presented in the previous sections, we have been able to obtain two-fluid solutions that are physically viable in the sense of the constraints given in \cite{del} and in Sec. \ref{CON}. 

A somewhat surprising but nonetheless interesting result concerns the generating theorems that have been shown to hold in the case of single fluid solutions. We discovered that one of these theorems (theorem 2 of \cite{boon1}) can also be employed to construct two-fluid solutions. We used this theorem to obtain a solution representing a two-fluid relativistic star comprising a perfect fluid and a fluid with a non trivial equation of state. 

As mentioned above, the equations we have constructed contain a complete description of the properties of the fluid, including fluxes. Since one can integrate in general the equation for the heat flux, for every two-fluid solution known, there is an additional one in which the fluxes are non-zero. We have given the expressions of the flux for all the solutions obtained in the text. We found that in Schwarzschild coordinates these fluxes are always divergent in the center of the matter distribution. As such divergence could be an artifact of the coordinates used for describing the metric, we do not consider the divergence of the flux as a reason to discard solutions as we have done with other matter potentials. Further investigation on this matter might shed light on the real nature of this feature. 

In general, the majority of the solutions we have found present a shell structure. In particular, we found that these solutions have to be completed with an additional shell which can be joined smoothly to any external spacetime we choose to consider (e.g., the Schwarzschild \cite{1916skpa.conf.424S}, Vaidya \cite{vaidya}, etc.). Such a shell can contain any number of fluids/fields consistent with the junction conditions. This should not be surprising: such configurations are expected to arise in multi-fluid systems, and the constraints arising from the junction conditions make such occurrences even more likely, as it is more complex to find solutions in which the pressures of two different fluids vanish at the same point.

The set of results presented above indicate clearly that the covariant formulation of the TOV equations is a powerful tool to investigate analytically the physics of interior solutions--even in the multi-fluid case. The possibility of obtaining physically reasonable and exact toy models allows one to explore more deeply the features of the interior of relativistic objects and their perturbations, and can be used as a complement to the large number of numerical models in literature as a testing tool.

We conclude by remarking that the value of our preliminary results goes beyond the scope of standard multi-fluid relativistic objects. The two-fluid approach described in this paper is also well-suited to the study of relativistic objects in theories beyond that of General Relativity, where the extra gravitational degree of freedom can be treated as an additional (curvature) fluid coupled to standard matter. This exciting application of our approach will be important in the context of gravitational wave astronomy, where for the first time the validity of General Relativity can be probed in the strong-field regime in regions of high density and scalar curvature. These issues will be explored in future papers. 

\begin{acknowledgments}
NFN wishes to acknowledge funding from the National Research Foundation (Grant number: 116629)
\end{acknowledgments}
\appendix
\section{Main aspects of the 1+1+2 covariant approach for a single fluid}\label{AA}
We will present the main aspects of the theory in this section. For more details, the reader is referred to \cite{clarkbar}, \cite{clarkson} and \cite{betschart}. 

We define a time-like threading vector field $u^a$ associated to the observer's congruence with $u^a u_a = -1$, and a space-like vector $e_a$ with $e_a e^a = 1$. The $u^a$ and $e_a$ congruences describe a geometry defined by two projection tensors given by
\begin{eqnarray}
{h^a}_b &=& {g^a}_b + u^a u_b \hspace{0.2cm},\hspace{0.2cm} {h^a}_a = 3, \nonumber \\
{N_a}^b &=& {h_a}^b - e_a e^b = {g_a}^b + u_a u^b - e_a e^b \hspace{0.1cm},\hspace{0.1cm}{N^a}_a = 2,\hspace{0.6cm}
\end{eqnarray}
where ${h^a}_b$ represents the metric of the 3-spaces orthogonal to $u_a$, and ${N_a}^b$ represents the metric of the 2-spaces orthogonal to $u_a$ and $e_a$. Any tensorial object may now be split according to the above foliations \cite{sante1}. The covariant time derivative, orthogonally projected covariant derivative, hat-derivative and $\delta$-derivative are given by
\begin{eqnarray}
{\dot{X}{}^{a..b}}_{c..d} &\equiv& u^e {\nabla}_e {X^{a..b}}_{c..d}, \nonumber \\
D_e {X^{a..b}}_{c..d} &\equiv& {h^a}_f...{h^b}_g{h^p}_c... {h^q}_d {h^r}_e {\nabla}_r {X^{f..g}}_{p..q}, \nonumber \\
{\hat{X}{}_{a..b}}^{c..d} &\equiv& e^f D_f {X_{a..b}}^{c..d}, \nonumber \\
{\delta}_e {X_{a..b}}^{c..d} &\equiv& {N_a}^f...{N_b}^g {N_i}^c...{N_j}^d {N_e}^p D_p {X_{f..g}}^{i..j}. \hspace{0.3cm}
\end{eqnarray}
The kinematical variables are given by
\begin{subequations} 
\begin{align} 
& \mathcal{A} = e_a \dot{u}^a, \hspace{0.2cm}\hspace{0.2cm} \mathcal{A}^a = N_{ab} \dot{e}^b,  \\
& \Theta=D_a u^a, \hspace{0.2cm} \xi = \frac{1}{2} \varepsilon^{ab} {\delta}_a e_b,  \\
& \Omega = \frac{1}{2} \varepsilon^{abc} D_{[a} u_{b]} e_a \hspace{0.2cm},\hspace{0.2cm}{\Omega}^a = \frac{1}{2} \varepsilon^{abd} D_{[a} u_{b]} {N_d}^a,  \\
& {\sigma}_{ab} = \left({n^c}_{(a} {h_{b)}}^d - \frac{1}{3} h_{ab} h^{cd} \right)D_c u_d,  \\
& \Sigma = {\sigma}^{ab} \left(e_a e_b - \frac{1}{2} N_{ab} \right) \hspace{0.2cm},\hspace{0.2cm} {\Sigma}_a = {\sigma}_{cd} e^c {N^a}_d, \\
& {\Sigma}_{ab} = \left({N^c}_{(a}{N_{b)}}^d - \frac{1}{2} N_{ab} N^{cd} \right){\sigma}_{cd},  \\
& a_b = e^c D_c e_b = \hat{e}_b \hspace{0.2cm},\hspace{0.2cm} \phi = {\delta}_a e^a,  \\
& {\zeta}_{ab} = \left({N^c}_{(a}{N_{b)}}^d - \frac{1}{2} N_{ab} N^{cd} \right){\delta}_c e_d,  \\
& \mathcal{E} = {C^{ac}}_{bd} u_c u^d \left(e_a e^b - \frac{1}{2} N_{a}^b \right),  \\
& \mathcal{E}_a = C_{cedf} u^e u^f e^c {N^d}_a \hspace{0.2cm},\hspace{0.2cm} \mathcal{E}_{ab} = C_{\{a}{}^{c}{}_{b\}}{}^d u_c u_d,  \\
& \mathcal{H} = \frac{1}{2} {\varepsilon^a}_{de} {C^{deb}}_c u^c \left(e_a e_b - \frac{1}{2} N_{ab}\right),  \\
& \mathcal{H}_a = \frac{1}{2}\varepsilon_{cfe} {C^{fe}}_{dh} u^h e^c {N^d}_a,  \\
& \mathcal{H}_{ab} = \frac{1}{2}{\varepsilon_{\{a}}^{de} C_{b\}cde} u^c,
\end{align}
\end{subequations} 
where $\varepsilon_{ab} \equiv \varepsilon_{abc} e^c$ and $\varepsilon_{abc} = \eta_{dabc} e^c u^d$ are the volumes of the two hypersurfaces, $C_{abcd}$ is the Weyl tensor. We represent the symmetrisation over the indices of a tensor as $T_{(ab)} = \frac{1}{2} (T_{ab} + T_{ba})$, and the anti-symmetrisation as $T_{[ab]} = \frac{1}{2} (T_{ab} - T_{ba})$. We use curly brackets $\{ \}$ to denote the Projected Symmetric Trace-Free with respect to $n^a$ part of a tensor:
\begin{equation}
X^{\{ab\}} 
\equiv \left( N^{c}{}_{(a}N_{b)}{}^{d} - \frac{1}{2}N_{ab} N^{cd}\right) \mathds{X}_{cd}~.
\end{equation}
The energy-momentum tensor is decomposed as
\begin{eqnarray}
T_{ab} = &&\mu u_a u_b + (p + \Pi) e_a e_b + \left(p - \frac{1}{2} \Pi \right) N_{ab} \nonumber\\
&& + 2 Q e_{(a} u_{b)} + 2 Q_{(a} u_{b)} + 2 \Pi_{(a} e_{b)} + \Pi_{ab}.
\end{eqnarray}
The matter variables are given by
\begin{eqnarray}
\mu &=& T_{ab} u^a u^b, \nonumber \\
p &=& \frac{1}{3} T_{ab} \left(e^a e^b + N^{ab}\right), \nonumber \\
\Pi &=& \frac{1}{3} T_{ab} \left(2 e^a e^b - N^{ab}\right), \nonumber \\
Q &=& \frac{1}{2} T_{ab} e^a u^b, \nonumber \\
Q_a &=& T_{cd} {N^c}_a u^d, \nonumber \\
\Pi_a &=& T_{cd} {N^c}_a e^d, \nonumber \\
\Pi_{ab} &=& T_{\{ab\}},
\end{eqnarray}
where $\mu$ is the density, $p$ is the pressure, $Q$ and $Q_a$ represent the scalar and vector parts of the heat flux, and $\Pi$ and $\Pi^a$ represents the the scalar and vector components of the anisotropic pressure.
The 1+1+2 formalism is most advantageous when applied to spacetimes which have a unique preferred spatial direction at each point, exhibiting local rotational symmetry (LRS). This direction constitutes a local axis of symmetry. In LRSII spacetimes, we have that the vorticity terms and the magnetic part of the Weyl tensor are zero, i.e., $\Omega, \xi, \mathcal{H}=0$. For static and spherically symmetric LRSII spacetimes, we also have $ \Theta, \Sigma=0$ and all the dot derivatives vanish.
The remaining 1+1+2 scalars which fully describe the spacetime can be divided into 3 categories \cite{betschart}:
\begin{subequations} \label{hatsystem1}
\begin{align} 
\intertext{\it Propagation:} \nonumber \\
\hat{\phi} &= -\frac{1}{2} {\phi}^2 - \frac{2}{3} \mu - \mathcal{E} - \frac{1}{2} \Pi, \\
Q &= 0, \\ 
\hat{\mathcal{E}} - \frac{1}{3} \hat{\mu} + \frac{1}{2} \hat{\Pi} &= -\frac{3}{2} {\phi} \left(\mathcal{E} + \frac{1}{2} \Pi \right). \\ 
\intertext{\it Evolution:} \nonumber \\
0 &= -\mathcal{A} \phi + \frac{1}{3}(\mu + 3p)- \mathcal{E} + \frac{1}{2} \Pi, \\
0&= \frac{1}{2} \phi Q . \\
\intertext{\it Propagation/evolution:} \nonumber \\
\hat{\mathcal{A}} &= - \mathcal{A} \left( \mathcal{A} + \phi \right) + \frac{1}{2}(\mu + 3p), \\
\hat{Q} &= -Q \left(\phi + 2 \mathcal{A} \right) + j_u, \\
\hat{p} + \hat{\Pi} &= -\Pi \left(\frac{3}{2} \phi + \mathcal{A} \right) - \mathcal{A} \left( \mu + p \right) + j_e,\\
K &= \frac{1}{3} \mu - \mathcal{E} - \frac{1}{2} \Pi + \frac{1}{4} {\phi}^2, \\ \nonumber
\end{align}
\end{subequations}
where $\mu$, $p$, $Q$ and $\pi$ represent, in general, the total energy density, pressure, total heat flux and total anisotropic pressure of the fluid. In addition, we have formally included the total particle interaction currents as $j_u$ and $j_e$ according to the definition
\begin{equation} \label{j}
j_{a} = j_u u_a + j_e e_a,
\end{equation}
with $j_u$ and $j_e$ representing the $u_a$ and $e_a$ components respectively.

Finally, it is not too difficult to prove that the Gauss curvature $K$ satisfies the propagation equation
\begin{equation} 
\hat{K}=-\phi K.
\end{equation}
\section{Multifluid equations}\label{App2}

The 1+1+2 equations in the case of a static spherically symmetric spacetime and $N$ different interacting fluids can be written as 
\begin{subequations} 
\begin{align} 
\hat{\phi} &=-\frac{1}{2} {\phi}^2 - \frac{2}{3} \sum_{i=1}^{N}{\mu}_i- \mathcal{E} - \frac{1}{2} \sum_{i=1}^{N}{\Pi}_i, \\
\sum_{i=1}^{N}Q_i &= 0, \\ 
\hat{\mathcal{E}} + \frac{3}{2} {\phi} \mathcal{E} &= \frac{1}{3} \sum_{i=1}^{N}{\hat{\mu}}_i - \frac{1}{2} \sum_{i=1}^{N}{\hat{\Pi}}_i- \frac{3}{2} {\phi} \frac{1}{2} \sum_{i=1}^{N}{\Pi}_i, \\
\mathcal{E} + \mathcal{A} \phi&= \frac{1}{3}\sum_{i=1}^{N}({\mu}_i+3p_i) + \frac{1}{2} \sum_{i=1}^{N}{\Pi}_i,\\
K&=\frac{1}{4} {\phi}^2- \mathcal{E} +\frac{1}{3} \sum_{i=1}^{N}{\mu}_i- \frac{1}{2} \sum_{i=1}^{N}{\Pi}_i,\\
\hat{\mathcal{A}}&=- \mathcal{A} \left( \mathcal{A} + \phi \right) + \frac{1}{2} \sum_{i=1}^{N}({\mu}_i+3p_i), \\
 \hat{Q}_i&=-Q_i \left(\phi + 2 \mathcal{A} \right) +\sum_{k\neq i }^{N}j_u^{(i,k)},\label{Qi}\\
\hat{p}_i + \hat{\Pi}_i&=\sum_{k\neq i }^{N}j_e^{(i,k)} -{\Pi}_i \left(\frac{3}{2} \phi + \mathcal{A} \right) - \mathcal{A} \left( {\mu}_i + p_i\right), \label{Pi}
\end{align}
\end{subequations}
where the index $i$ represents the i-th component and $j_e^{(i,k)}$ is the interaction term between the component $i$ and $k$. Notice that $j_u^{(i,j)}=-j_u^{(j,i)}$ and $j_e^{(i,j)}=-j_e^{(j,i)}$.

By summing the \eqref{Pi} over $i$ we have the equation for the total energy pressure : 
\begin{align}
\hat{p}_{tot} + \hat{\Pi}_{tot} &= -\sum_{i=1}^{N}\Pi_1 \left(\frac{3}{2} \phi + \mathcal{A} \right) - \mathcal{A} \sum_{i=1}^{N}\left( {\mu}_i + p_i \right).
\end{align}
In terms of the variable $\rho$ defined by the relation $K =  {K_0}^{-1} e^{- \rho},$ the 1+1+2 equations can be written as
\begin{subequations}
\begin{align} 
\phi {\phi}_{, \rho} &=-\frac{1}{2} {\phi}^2 - \frac{2}{3} \sum_{i=1}^{N}{\mu}_i- \mathcal{E} - \frac{1}{2} \sum_{i=1}^{N}{\Pi}_i, \\
\sum_{i=1}^{N}Q_i &= 0, \\
{\mathcal{E}}_{, \rho} + \frac{3}{2} \mathcal{E} &= \frac{1}{3} \sum_{i=1}^{N}{{\mu}}_{i, \rho} - \frac{1}{2} \sum_{i=1}^{N}{{\Pi}}_{i, \rho} - \frac{3}{2} {\phi} \frac{1}{2} \sum_{i=1}^{N}{\Pi}_i,\\
\mathcal{E} + \mathcal{A} \phi&= \frac{1}{3}\sum_{i=1}^{N}({\mu}_i+3p_i) + \frac{1}{2} \sum_{i=1}^{N}{\Pi}_i,
\end{align}
\begin{align} 
K&=\frac{1}{4} {\phi}^2- \mathcal{E} +\frac{1}{3} \sum_{i=1}^{N}{\mu}_i- \frac{1}{2} \sum_{i=1}^{N}{\Pi}_i,\\
\phi{\mathcal{A}}_{, \rho}&=- \mathcal{A} \left( \mathcal{A} + \phi \right) + \frac{1}{2} \sum_{i=1}^{N}({\mu}_i+3p_i), \\
\phi \hat{Q}_{i, \rho}&=-Q_i \left(\phi + 2 \mathcal{A} \right) +\sum_{k\neq i }^{N}j_u^{(i,k)},\\
\hat{p}_i + \hat{\Pi}_i&=\sum_{k\neq i }^{N}j_e^{(i,k)} -{\Pi}_i \left(\frac{3}{2} \phi + \mathcal{A} \right) - \mathcal{A} \left( {\mu}_i + p_i\right).
\end{align}
\end{subequations}
Using the variables 
\begin{align}
X &= \frac{{\phi}_{, \rho}}{\phi}, & Y &= \frac{\mathcal{A}}{\phi}, & \mathcal{K} &= \frac{K}{{\phi}^2} ,\nonumber\\
E &= \frac{\mathcal{E}}{{\phi}^2}, & \mathbb{M}_1 &= \frac{{\mu}_1}{{\phi}^2}, &
\mathbb{M}_2 &= \frac{{\mu}_2}{{\phi}^2}, \nonumber\\
P_1 &= \frac{p_1}{{\phi}^2}, & P_2 &= \frac{p_2}{{\phi}^2}, &
\mathbb{P}_1 &= \frac{{\Pi}_1}{{\phi}^2} ,\label{newvarsb}\\
\mathbb{P}_2 &= \frac{{\Pi}_2}{{\phi}^2} ,& \mathbb{Q}_1 &= \frac{Q_1}{{\phi}^2}, & 
\mathbb{Q}_2 &= \frac{Q_2}{{\phi}^2} ,\nonumber\\ 
\mathbb{J}_u &= \frac{j_u^{(1,2)}}{{\phi}^3}, & \mathbb{J}_e &= \frac{j_e^{(1,2)}}{{\phi}^3},\nonumber
\end{align}
we obtain the covariant TOV equations as
\begin{align}\label{tovN} 
{P}_{i, \rho} + \mathbb{P}_{i, \rho}  =& \sum_{i\neq k}^N\mathbb{J}_e^{(i,k)}- {P_i}^2 - {\mathbb{P}_i}^2 \nonumber \\
&+ P_i \left[\mathbb{M}_i - 2\mathbb{P}_i - 3 \mathcal{K} + \frac{7}{4} \right] \nonumber \\
&+\mathbb{P}_i \left(\mathbb{M}_i - 3\mathcal{K} + \frac{1}{4}\right) \nonumber \\
& + \mathbb{M}_i \left(\frac{1}{4} -\mathcal{K}\right) \nonumber 
\end{align} 
\begin{align} 
& - P_i\sum_{k\neq i }^{N} (P_k + \mathbb{P}_k - 2\mathbb{M}_k) \nonumber \\
&- \mathbb{P}_i\sum_{k\neq i }^{N}(P_k + \mathbb{P}_k - 2\mathbb{M}_k) \nonumber \\
&- \mathbb{M}_i\sum_{k\neq i }^{N}(P_k + \mathbb{P}_k), \\
\mathcal{K}_{, \rho} =& 2 \mathcal{K} \left( \frac{1}{4}-\mathcal{K} + \sum_{k= 1 }^N\mathbb{M}_k \right), \\
\mathbb{Q}_{i, \rho}=& \mathbb{Q}_i \left[ 2\mathcal{K} - 2 \sum_{k= 1 }^N\mathbb{M}_k-\frac{3}{2}\right]\nonumber\\
&- \sum_{i\neq k}^N\mathbb{J}_u^{(i,k)},
\end{align}
where we have set
\begin{equation}
\mathbb{J}_u^{(i,k)}=\frac{j_u^{(i,k)}}{{\phi}^3},\quad \mathbb{J}_e^{(i,k)}= \frac{j_e^{(i,k)}}{{\phi}^3}.
\end{equation}
In the case of two fluids the equations above reduce to
\begin{subequations} 
\begin{align} 
\hat{\phi}=&-\frac{1}{2} {\phi}^2 - \frac{2}{3} ({\mu}_1 + {\mu}_2) - \mathcal{E} \nonumber\\
&- \frac{1}{2} ({\Pi}_1 + {\Pi}_2), \\
\hat{\mathcal{E}}=& \frac{1}{3} (\hat{\mu}_1 - \hat{\mu}_2) - \frac{1}{2} ( \hat{\Pi}_1 + \hat{\Pi}_2)\nonumber\\
&-\frac{3}{2} {\phi} \left(\mathcal{E} + \frac{1}{2} {\Pi}_1 + \frac{1}{2} {\Pi}_2 \right), \\ 
-\mathcal{A} \phi &+ \frac{1}{3}({\mu}_1 + 3 p_1) + \frac{1}{3}({\mu}_2 + 3 p_2)\nonumber \\ 
&- \mathcal{E} + \frac{1}{2} {\Pi}_1 + \frac{1}{2} {\Pi}_2 = 0,\\
\hat{\mathcal{A}}=&- \mathcal{A} \left( \mathcal{A} + \phi \right) + \frac{1}{2}({\mu}_1 + 3p_1) \nonumber\\ 
&+ \frac{1}{2}({\mu}_2 + 3p_2),\\
 \hat{p}_1 + \hat{\Pi}_1=&-{\Pi}_1 \left(\frac{3}{2} \phi + \mathcal{A} \right) - \mathcal{A} \left( {\mu}_1 + p_1 \right) + j_e^{(1,2)}, \\ 
 \hat{p}_2 + \hat{\Pi}_2 =&-{\Pi}_2 \left(\frac{3}{2} \phi + \mathcal{A} \right) - \mathcal{A} \left( {\mu}_2 + p_2 \right) - j_e ^{(2,1)},\\
K=&\frac{1}{3} ({\mu}_1 + {\mu}_2) - \mathcal{E} - \frac{1}{2} ({\Pi}_1 + {\Pi}_2) + \frac{1}{4} {\phi}^2 ,\\
\hat{Q}_1=&-Q_1 \left(\phi + 2 \mathcal{A} \right) + j_u^{(1,2)}, \\
Q_2 =& - Q_1,\\ 
 j_u^{(1,2)}&=- j_u^{(2,1)}\\
 j_e^{(1,2)}&=- j_e^{(2,1)}
\end{align}
\end{subequations}
which are given in \eqref{hatsystem2} with $\Pi=0$, $ j_u^{(1,2)}=0$ and $ j_e^{(1,2)}=0$. In terms of $\rho$
 \begin{subequations}
\begin{align} 
\phi{\phi}_{,\rho}=&-\frac{1}{2} {\phi}^2 - \frac{2}{3} ({\mu}_1 + {\mu}_2) - \mathcal{E} \nonumber\\
&- \frac{1}{2} ({\Pi}_1 + {\Pi}_2), \\
{\mathcal{E}}_{, \rho }=& \frac{1}{3} ({\mu}_1 - {\mu}_2) - \frac{1}{2} ({\Pi}_1 + {\Pi}_2)\nonumber \\  
&-\frac{3}{2} \left(\mathcal{E} + \frac{1}{2} {\Pi}_1 + \frac{1}{2} {\Pi}_2 \right), 
\end{align}
\begin{align} 
-\mathcal{A} \phi &+ \frac{1}{3}({\mu}_1 + 3 p_1) + \frac{1}{3}({\mu}_2 + 3 p_2) \nonumber \\ 
&- \mathcal{E} + \frac{1}{2} {\Pi}_1 + \frac{1}{2} {\Pi}_2 = 0,\\
\phi{\mathcal{A}}_{, \rho }=&- \mathcal{A} \left( \mathcal{A} + \phi \right) + \frac{1}{2}({\mu}_1 + 3p_1) \nonumber\\
&+ \frac{1}{2}({\mu}_2 + 3p_2), \\
 \phi\left({p}_{1, \rho }+ {\Pi}_{1, \rho }\right)=&-{\Pi}_1 \left(\frac{3}{2} \phi + \mathcal{A} \right) \nonumber 
\end{align} 
\begin{align} 
 &- \mathcal{A} \left( {\mu}_1 + p_1 \right) + j_e^{(1,2)}, \\
 \phi\left({p}_{2, \rho }+ {\Pi}_{2, \rho }\right)=&-{\Pi}_2 \left(\frac{3}{2} \phi + \mathcal{A} \right) \nonumber \\
 &- \mathcal{A} \left( {\mu}_2 + p_2 \right) - j_e^{(1,2)} ,\\
K+ \mathcal{E}-\frac{1}{4} {\phi}^2 =&\frac{1}{3} ({\mu}_1 + {\mu}_2) - \frac{1}{2} ({\Pi}_1 + {\Pi}_2) ,\\
\phi{Q}_{1, \rho}=&-Q_1 \left(\phi + 2 \mathcal{A} \right) + j_u^{(1,2)}, \\
Q_2 =& - Q_1, 
\end{align}
\end{subequations}
and the corresponding TOV equations, for $\Pi=0$, are given in \eqref{tov2}.

\end{document}